\documentclass[journal]{IEEEtran}

\usepackage{epsf}
\usepackage{float}
\usepackage{graphicx}
\DeclareGraphicsExtensions{.png}
\usepackage{caption}
\usepackage{subcaption}
\usepackage{xcolor}
\usepackage{amsmath}\interdisplaylinepenalty=2500
\usepackage{amsfonts}
\usepackage{amssymb}
\usepackage{array}
\usepackage{cuted}

\newcommand{\ul}[1]{\underline{#1}}

\newcommand{\sinc}{{\rm sinc}}

\newcommand{\bea}{\begin{eqnarray}}

\newcommand{\eea}{\end{eqnarray}}

\newcommand{\rmd}{{\rm d}}

\newcommand{\rme}{{\rm e}}

\newcommand{\rmj}{{\rm j}}

\newcommand{\cA}{{\cal A}}

\newcommand{\cB}{{\cal B}}

\newcommand{\cD}{{\cal D}}

\newcommand{\cE}{{\cal E}}

\newcommand{\cF}{{\cal F}}

\begin{document}



\title{Correlation and Spectral Density Functions in Mode-Stirred Reverberation -- 
II. Spectral Moments, Sampling, Noise, EMI and Understirring
}

\author{
{Luk R. Arnaut 
and John M. Ladbury
}\\
}


\maketitle



\begin{abstract} 
In part I, spectral moments and kurtosis were established as parameters in analytic models of correlation and spectral density functions for dynamic reverberation fields. In this part II, several practical limitations affecting the accuracy of estimating these parameters from measured stir sweep data are investigated. For sampled fields, the contributions of finite differencing and aliasing are evaluated. Finite differencing results in a negative bias that depends, to leading order, quadratically on the product of the sampling time interval and the stir bandwidth. Numerical estimates of moments extracted directly from sampled stir sweeps show good agreement with values obtained by an autocovariance method. The effects of data decimation and noise-to-stir ratios of RMS amplitudes are determined and experimentally verified. In addition, the dependencies on the noise-to-stir-bandwidth ratio, EMI, and unstirred energy are characterized.
\end{abstract}


{\bf \small {\it Index Terms}--spectral moments, stir noise, stir spectral density.}

\section{Introduction}
In part I \cite{arnaACFSDF_pt1}, analytical models of correlation functions (CFs) and spectral density functions (SDFs) for continuously stirred fields and power inside a mode-stirred reverberation chamber (MSRC) were developed. 
The normalized second spectral moment (spectral variance) and the scaled spectral kurtosis were found to be their key parameters that require estimation and evaluation.
In this part II, the influence of various properties of the physical stirred field (noise level and bandwidth, understirred field components, external EMI) and several features in the data acquisition (sample size, sampling rate, stir speed, IF bandwidth (IFBW)) on the estimation of spectral moments and kurtosis is analyzed. The primary aim is to quantify and compensate for bias resulting from these effects in experimental results for CFs and SDFs presented in part III \cite{arnaACFSDF_pt3}. 
The analysis is confined to linear effects of nonlocal temporal second order (two-point stir correlation) and local spectral fourth order (stir spectral kurtosis). Sampling effects on local temporal first-order statistics and probability density functions (PDFs) have been widely studied previously, e.g., \cite{arnaTQE2}, \cite{west2012}. 
As shown in \cite{arnaACFSDF_pt3}, discrete autocorrelation functions (ACFs) and autospectral density functions (ASDFs) of MSRC data can be applied in performing stir diagnostics.

Specifically, the effect of finite differencing (FD) on the bias and variance of spectral moments is analyzed in detail. Their evaluation based on the sweep data   
is compared with that from the autocovariance (ACV) method, 
showing similar levels of performance while being conceptually simpler. 
Whereas the methodology applies to general ACFs and SDFs, explicit expressions are derived for the exponential ASDF in \cite{arnaACFSDF_pt1}.

Theoretical results are evaluated and illustrated with measured data obtained using a vector network analyzer (VNA). Notwithstanding the high sampling rate of modern VNAs enabling near-continuous acquisition of stir process data (i.e., stir traces as sample paths), the implication of discretization remains an inherent issue for continuous random fields of unknown bandwidth. Conversely, sampling allows for approaching and investigating fundamental limits and trade-offs between accuracy, uncertainty and duration of measurement, as will be shown.

The following conventions and notations are adopted. $m$ overdots indicate $m$th-order continuous differentiation or discrete differencing. Single- and double-primed quantities relate to the real (in-phase, I-) and imaginary (quadrature, Q-) parts of an assumed wide-sense stationary (WSS) complex random electric field $E(\tau) = E^\prime(\tau) + \rmj E^{\prime\prime}(\tau)$, to its complex CF $\rho_E(\tau)=\rho^\prime_E(\varpi)+\rmj \rho^{\prime\prime}_E(\varpi)$ at stir lag $\tau$, or to its single-sided SDF $g_E(\varpi) = g^\prime_E(\varpi)+\rmj g^{\prime\prime}_E(\varpi)$ at stir frequency $\varpi=2\pi/\tau$. 
Spectral moments are denoted as $\lambda_m$ or $\lambda^{\prime(\prime)}_m$, the latter implying a normalization by $\lambda_0 = \sigma^2_E$ and association with $g^{\prime(\prime)}_E(\varpi)$. For other notations, cf. \cite{arnaACFSDF_pt1}.

\section{Spectral Moments of Sampled Fields\label{sec:FDdef}}
Spectral-statical properties of $E(\tau)$ can be established by exploiting the duality between the PDF and the characteristic function of $E(\tau)$ as a pair of Fourier transforms on one hand, versus its SDF and CF as another such pair on the other hand.  
A characterization of CFs and SDFs can be based on spectral moments.
The spectral moments $\lambda_m$ of the normalized single-sided analytic SDF $g_E(\varpi\geq 0)$ are related to the averaged product of $n$th- and $p$th-order stir time derivatives of $E^{(*)}(\tau)$ as \cite{long1956}--\cite{vanm1983} 
\begin{align}
&\lambda_{m} 
\stackrel{\Delta}{=} 
\sigma^2_{E} \int^{+\infty}_0 \varpi^m g_{E}(\varpi) \rmd \varpi =
\sigma^2_E \int^{+\infty}_{-\infty} |\varpi|^m s_E(\varpi) \rmd \varpi
\label{eq:def_lambdam_derivatives_temp}\\
&= \rmj^{-n} (-\rmj)^{-p} 
\overline{E^{(n)}(\tau)
{E^*}^{(p)}(\tau) },\hspace{2mm}
~n+p=m=0,1,2,\ldots
\label{eq:def_lambdam_derivatives}
\end{align}
where $s_E(|\varpi|) = g_{E}(\varpi\geq 0) / 2$ represents the normalized double-sided SDF.
In the current framework, 
the overline in (\ref{eq:def_lambdam_derivatives}) denotes sample path averaging with respect to $\tau$, i.e., taken across a single sweep of the primary mode stirrer, for an arbitrary state of the secondary tuner, $\tau_{2,\ell}$. For arbitrary order $m$, the $\lambda_m$ are sample values of random $\Lambda_m$
with PDF $f_{\Lambda_m}(\lambda_m)$.
Correspondingly, $g_E(\varpi)$ is a sample SDF from an ensemble of SDFs generated by secondary tuning.
The focus is on effective `macroscopic' SDFs of the observable interior cavity field. 
In principle, a microscopic characterization \cite{vanm1972} based on moments of SDFs for individual cavity eigenmodes and their stir evolution is also possible.

Note that (\ref{eq:def_lambdam_derivatives}) is ambiguous, in that different combinations of $n$ and $p$ values may yield the same $m$ but not necessarily the same value of $\lambda_m$.
This ambiguity increases with increasing $m$, as more combinations of $n$ and $p$ become valid. This could be resolved by redefining $\lambda_m$ as the arithmetic mean 
\begin{align}
\frac{\rmj^m}{m+1} \sum^m_{n=0} (-1)^n
\overline{
E^{(n)}(\tau) {E^*}^{(m-n)}(\tau)
}
.
\end{align}

On the other hand, FD of sampled data provides approximations to the sought continuous derivatives in (\ref{eq:def_lambdam_derivatives}), {\em a fortiori} for large $m$.
As clarified in sec. \ref{sec:eff_finitediff}, only combination(s) of $n$ and $p$ that minimize(s) $|n-p|$  will be retained, i.e., $n=p=m/2$ for $m$ even and $n=p-1=\lfloor m/2 \rfloor$ for $m$ odd. 
This choice yields close numerical agreement with ACV-based estimates of $\lambda_m$.
Thus, (\ref{eq:def_lambdam_derivatives}) will be evaluated as
\begin{align}
\lambda_{2i} &= 
\overline{
\left [ {E^\prime}^{(i)}(\tau) \right ]^2 
}
+
\overline{ 
\left [ {E^{\prime\prime}}^{(i)}(\tau) \right ]^2
}
\rightarrow 
2 
\overline{\left [
{E^{\prime(\prime)}}^{(i)}(\tau) \right ]^2 
}
\label{eq:lambda_2i_nequalp}
\\
\lambda_{2i+1} &= 
\overline{
{E^\prime}^{(i)}(\tau) 
{E^{\prime\prime}}^{(i+1)}(\tau) 
}
-
\overline{
{E^{\prime}}^{(i+1)}(\tau) 
{E^{\prime\prime}}^{(i)}(\tau) 
}
\nonumber\\
&\hspace{1cm} \rightarrow
2 
\overline{ {E^\prime}^{(i)}(\tau) 
{E^{\prime\prime}}^{(i+1)}(\tau) 
}\label{eq:lambda_2iplus1_nequalpmin1}
\end{align}
for $m$ even or odd , respectively.
In (\ref{eq:lambda_2i_nequalp}) and (\ref{eq:lambda_2iplus1_nequalpmin1})
 the limit expressions on the right
hold for ideal circular $E(\tau)$, for which 
$\overline{{E^\prime}^{(m-n)}(\tau) {E^\prime}^{(n)}(\tau)} = \overline{{E^{\prime\prime}}^{(n)}(\tau)
{E^{\prime\prime}}^{(m-n)}(\tau)}$
 and 
$\overline{{E^\prime}^{(m-n)}(\tau)
{E^{\prime\prime}}^{(n)}(\tau)} = 
- \overline{{E^\prime}^{(n)}(\tau)
{E^{\prime\prime}}^{(m-n)}(\tau)}$.

For consistency of notation, we shall further replace the overbar for temporal averaging across a stir sweep with $\langle \cdot \rangle_{N_s}$ (or simply $\langle \cdot \rangle$, at an arbitrary tune state, if there is no confusion), while reserving $\langle \cdot \rangle_{N_t}$ to represent ensemble averaging over secondary tune states at an arbitrary stir state.

In the literature on spectral moments, e.g.,  
\cite{mill1970b}--\cite{meln2004}, 
the main focus has been on the effect of finite sample size on the bias and variance in the estimation of $\lambda_m$, and on comparing temporal vs. spectral methods of estimation, restricted almost exclusively to low orders ($m=0,1,2$) and  Gaussian SDFs. 
Results on higher-order moments and non-Gaussian SDFs are relatively scant \cite{mona1999}.
Here, we compare the bias caused by FD of continuous vs. sampled stir data, for arbitrary $m$. General results are then applied to the $[0/1]$-order exponential SDF \cite[eq. (34)]{arnaACFSDF_pt1}, for which
\begin{align}
\lambda^{\prime(\prime)}_{m} 
&= {m!}/{(\beta^{\prime(\prime)})^{m}},\hspace{0.5cm}m=0,1,2\ldots 
\label{eq:specmom_ideal1storderPade}
\end{align}
 
\section{Bias Caused by Finite Differencing \label{sec:FDbias}}
\subsection{Estimation Based on Stir Sweep Data\label{subsec:FDbiasSweep}}

\subsubsection{Continuous Stir Sweep\label{sec:contstirsweep}}
The approximation of an $m$th-order continuous-time derivative $E^{(m)}(\tau)$ through FD can be represented using a scaled $m$-fold convolution of $E(\tau)$ with the anti-symmetric impulse pair function 
 \cite{brac1986}
\begin{align}
\Pi_{\Delta \tau}(\tau) \stackrel{\Delta}{=} \delta(\tau+\Delta\tau/2) - \delta(\tau-\Delta\tau/2).
\end{align} 
In particular, the first-order FD of $E(\tau)$ is
\begin{align}
{\cD}^1_{\Delta\tau} [E(\tau)] &\stackrel{\Delta}{=} \frac{E(\tau+\Delta \tau/2)-E(\tau-\Delta\tau/2)}{\Delta\tau} \label{eq:def_firstorderdiff_stirsweep}\\
&= |\Delta\tau|^{-1} \Pi_{\Delta \tau}(\tau) * E(\tau)
\label{eq:def_firstorderdiff_stirsweep_conv}
\end{align}
where $*$ denotes linear convolution. Since
\begin{align}
\cF [|\Delta\tau|^{-1} \Pi_{\Delta\tau}(\tau)](\varpi) = \rmj (2/\Delta \tau) \sin(\varpi \Delta\tau / 2)
\label{eq:cft_firstorderdiff_stirsweep_conv}
\end{align}
it follows that the first-order FD corresponds to the multiplication of ${\cE}(\varpi)\stackrel{\Delta}{=}\cF[E(\tau)](\varpi)$ by $\rmj \varpi \, \sinc(\varpi\Delta\tau/2)$. More generally, for $m$th-order FD involving $m$-fold convolution
\begin{align}
\cF[{\cD}^m_{\Delta \tau} [ E(\tau) ] ](\varpi) = [\rmj \varpi\, {\sinc(\varpi \Delta\tau / 2)}]^m\, {\cE}(\varpi)
.
\label{eq:cft_mthorderdiff_stirsweep_conv}
\end{align}
 
Parenthetically, a formal correspondence exists between discrete FD (\ref{eq:def_firstorderdiff_stirsweep})--(\ref{eq:cft_mthorderdiff_stirsweep_conv}) and continuous local averaging, i.e.,
\begin{align}
{\cA}^1_{\Delta\tau} [E(\tau)] &\stackrel{\Delta}{=} |{\Delta\tau}|^{-1} \int^{\Delta\tau/2}_{-\Delta\tau/2} E(\tau+s) \rmd s 
\label{eq:def_locavg_stirsweep}\\
\cF [{\cA}^m_{\Delta\tau} [ E(\tau) ] ](\varpi) &= [\sinc(\varpi \Delta\tau/2)]^m \cE(\varpi)
.
\label{eq:cft_mthorderlocavg_locavg_conv}
\end{align}
In particular, for uniform first-order averaging ($m=1$), several of the following methods and results for FD of $E(\tau)$ apply {\it mutatis mutandis} to local averaging of $E(\tau)$ \cite{arnalocavg}.  
 
Returning to (\ref{eq:cft_mthorderdiff_stirsweep_conv}), since 
$g^\prime_E(\varpi) = \langle \cE(\varpi)\cE^*(\varpi) \rangle / \sigma^2_E$, the FD of $E(\tau)$ corresponds to a multiplication of $g_E(\varpi)$ by $\sinc^{m}(\varpi \Delta\tau / 2)$.
Its effect on the spectral moments is
\begin{align}
\lambda^{\prime(\prime)}_{m,fd,c} 
&= \int^{+\infty}_0 \varpi^{m} \sinc^{m}\left ({\varpi \Delta\tau }/{2} \right ) g^{\prime(\prime)}_E(\varpi) \rmd \varpi
\label{eq:def_lambdaeven_discr}
\end{align}
where the subscript $fd,c$ denotes FD performed on a continuous function (analog stir sweep).

For a continuous and mean-square differentiable $E(\tau)$ \cite{wong1985},
its ASDF can be represented by the $[0/1]$ Pad\'{e} model 
$
\tilde{g}^\prime_E(\varpi) = \beta^\prime \exp(-\beta^\prime \varpi)
$ 
with $\beta^\prime = \sqrt{2/\lambda^\prime_{2}}$.
An independent or iterative estimate of $\beta^\prime$ is desirable. 
For even-order moments ($m=2i$), substituting this $\tilde{g}^\prime_E(\varpi)$ into (\ref{eq:def_lambdaeven_discr}) yields
\begin{align}
\lambda^\prime_{2i,fd,c} &=
\overline{\left |\frac{\Delta^i E(\tau)}{\Delta \tau^i} \right |^2}
= \frac{(2i )!}{(\beta^\prime)^{2i}} 
\prod^i_{k=1} \left [ 1 + \left ( \frac{k\Delta \tau}{\beta^\prime} \right )^2 \right ]^{-1}
\label{eq:estlambdapeven_cont_fd}
.
\end{align}
If $\Delta\tau / \beta^\prime \ll 1$ then, to leading order, (\ref{eq:estlambdapeven_cont_fd}) is approximated as
\begin{align}
\lambda^\prime_{2i,fd,c} 
&\simeq \frac{(2i) !}{(\beta^\prime)^{2i}} 
\left [ 1 - \frac{i(i+1)(2i+1)}{6} \left ( \frac{\Delta\tau}{\beta^\prime} \right )^2  \right ]
.
\label{eq:estlambdapeven_cont_fd_approx}
\end{align}
In particular, the estimated RMS power $\lambda_0$ is not affected by this FD on continous $E(\tau)$, as expected, while the relative bias of $\lambda^\prime_{2i}$ for $i>0$ increases rapidly with $i$.
Compensation for FD bias yields debiased moments as
\begin{align}
\lambda^\prime_{2i} &= \lambda^\prime_{2i,fd,c} 
\prod^i_{k=1} \left [ 1 + \left ( \frac{k\Delta \tau }{\beta^\prime} \right )^2 \right ] \\
&\simeq
\lambda^\prime_{2i,fd,c} \left [ 1 + \frac{i(i+1)(2i+1)}{6} \left ( \frac{\Delta\tau}{\beta^\prime} \right )^2  \right ]
.
\end{align}
The FD bias of $\kappa^\prime\equiv \lambda^\prime_{4}/]6(\lambda^\prime_{2})^2]-1$ follows as
\begin{align}
\Delta \kappa^\prime_{fd,c} &= \frac{1+\left (\Delta \tau/\beta^\prime \right )^2}{ 1+ 4 \left (\Delta \tau/\beta^\prime \right )^2}  - 1 \\
&\simeq
\left \{
\begin{array}{ll}
- 3 \left ( \Delta \tau/\beta^\prime \right )^2,&\Delta \tau/\beta^\prime \ll 1\\
-3/4,&\Delta \tau/\beta^\prime \gg 1
\end{array}
\right.
\label{eq:bias_kappap_fd_cont}
\end{align}
which must be similarly compensated for when estimating $\kappa^\prime$.

For odd-order moments ($m=2i+1$), the $[1/2]$-order model $\tilde{g}^{\prime\prime}_E(\varpi) = \beta^{\prime\prime} \exp(-\beta^{\prime\prime}\varpi)$  with $\beta^{\prime\prime}=\sqrt{6\lambda^{\prime\prime}_1/\lambda^{\prime\prime}_3}$ in (\ref{eq:def_lambdaeven_discr}) yields
\begin{align}
&\lambda^{\prime\prime}_{2i+1,fd,c} 
= \frac{(2i+1)!}{(\beta^{\prime\prime})^{2i+1}} 
\prod^i_{k=0} \left [ 
1 + 
    \left ( \frac{\left ( k + \frac{1}{2} \right ) \Delta \tau}{\beta^{\prime\prime}} 
    \right )^2 
\right ]^{-1}
\label{eq:def_lambdaodd_discr}
\\
&\simeq  
\frac{(2i+1)!}{(\beta^{\prime\prime})^{2i+1}}
\left [ 1 - \frac{4 i^3+12i^2+11i+3}{12} 
\left ( \frac{\Delta\tau}{\beta^{\prime\prime}} 
\right )^2  
\right ]
\label{eq:def_lambdaodd_discr_approx_explicit}
\end{align}
where the latter approximation again holds for $\Delta\tau / \beta^{\prime\prime} \ll 1$.
After compensation for FD bias, these become
\begin{align}
&\lambda^{\prime\prime}_{2i+1} = \lambda^{\prime\prime}_{2i+1,fd,c}
\prod^i_{k=0} \left [ 
1 + 
    \left ( \frac{\left ( k + \frac{1}{2} \right ) \Delta \tau}{\beta^{\prime\prime}} 
    \right )^2 
\right ]
\label{eq:def_lambdaodd_discr_comp}\\
&\simeq  
\lambda^{\prime\prime}_{2i+1,fd,c}
\left [ 1 + \frac{4 i^3+12i^2+11i+3}{12} 
\left ( \frac{\Delta\tau}{\beta^{\prime\prime}} 
\right )^2  
\right ]
\label{eq:def_lambdaodd_discr_approx_comp_explicit}
\end{align}
resulting in the FD debiased squared CSDF bandwidth
\begin{align}
\frac{\lambda^{\prime\prime}_{3}}{6\lambda^{\prime\prime}_{1}}
=
\frac{\lambda^{\prime\prime}_{3,fd,c}}{6\lambda^{\prime\prime}_{1,fd,c}}
\left [ 
1 + \frac{9}{4} \left ( \frac{\Delta \tau}{ \beta^{\prime\prime}} \right )^2 \right ]
.
\end{align}
The FD bias of $\kappa^{\prime\prime}\equiv 3 \lambda^{\prime\prime}_1\lambda^{\prime\prime}_5/[10(\lambda^{\prime\prime}_3)^2]-1$ follows as
\begin{align} 
\Delta\kappa^{\prime\prime}_{fd,c} &= 
\frac{ 1 + 9 \left ( {\Delta \tau} / {\beta^{\prime\prime}} \right )^2 / 4 }
{ 1 + 25 \left ( {\Delta \tau} / {\beta^{\prime\prime}} \right )^2 / 4 } - 1 \\
&\simeq
\left \{
\begin{array}{ll}
- 4 \left ( \Delta \tau/\beta^{\prime\prime} \right )^2,&\Delta \tau/\beta^{\prime\prime} \ll 1\\
-16/25,&\Delta \tau/\beta^{\prime\prime} \gg 1.
\end{array}
\right.
\label{eq:bias_kappapp_fd_cont}
\end{align}

In summary, FD of a continuous stir trace results in a negative quadratic bias proportional to $- (\Delta \tau/\beta^{\prime(\prime)})^2$ when $\Delta \tau/\beta^{\prime(\prime)} \ll 1$, to leading order. This bias underestimates all spectral moments and $\kappa^{\prime(\prime)}$. At higher CW frequencies (implying larger stir bandwidths $1/\beta^{\prime(\prime)}$) or for slower sampling, this bias increases.

\subsubsection{Sub-Nyquist Sampling}
When sampling a continuous stir sweep at $N_s$ uniformly spaced points $\tau_n \equiv n\Delta \tau$, the spectral expansion of $E(\tau)$
\cite[eq. (3)]{arnaACFSDF_pt1} 
in based on
the stir discrete-time Fourier transform (DTFT)
\begin{align}
{E} (\tau_n) 
&= \frac{\Delta\tau}{2\pi} \int^{+\pi/\Delta\tau}_{-\pi/\Delta\tau} {\cE}(\varpi) \exp (\rmj\varpi n\Delta\tau ) \rmd \varpi
.
\label{eq:expansion_tau_discr}
\end{align}
The corresponding stir IDFT at $\tau_n$ is a discrete sum over $N_s$ stir frequencies $\varpi_k = k\Delta \varpi$ ($k=0,1,\ldots, N_s-1)$ with stir-spectral resolution $\Delta \varpi = (2\pi/\Delta \tau)/N_s$, i.e.,
\begin{align}
{E} (n) 
&= \frac{1}{N_s} \sum^{N_s-1}_{k=0} \ul{\cE}(\varpi_k) \exp (\rmj 2 \pi k n / N_s) 
\label{eq:expansion_tau_discr_varpi_discr}
.
\end{align}
Therefore, for discretized stir sweeps, \cite[{\rm eq.} (33)]{arnaACFSDF_pt1} becomes
\begin{align}
\{ E(\tau_n) \} \stackrel{{\rm DFT}}{\longrightarrow} 
\{ \cE(\varpi_k) \} 
\stackrel{[1, (32)]}{\longrightarrow} 
\{ g_{E}(\varpi_k) \} \stackrel{{\rm IDFT}}{\longrightarrow} 
\{ \rho_E(\tau_n) \}
.
\label{eq:periodogram_discr}
\end{align}
The discrete SDF $g_E(\varpi_{\textcolor{blue}{k}})$ is obtained from either the DFT of the sampled field (periodogram based SDF) or its correlation matrix,
with the integration limits $\pm\infty$ replaced by the Nyquist limits, $\pm \varpi_{\rm max} = \pm \pi/\Delta\tau$.

The estimated spectral moments are now
\begin{align}
\lambda^{\prime(\prime)}_{m,fd,d} 
&= \alpha^{\prime(\prime)}_{m,fd,d} \lambda^{\prime(\prime)}_{m,fd,c} 
\label{eq:def_lambdaevenodd_discr_sampleddata}
\end{align}
where the subscript $fd,d$ refers to FD performed on discrete stir data, with \cite[eq. (3.895.3)]{grad2007}
\begin{align}
&\alpha^\prime_{2i,fd,d}
=
1 - \exp\left (- \frac{\pi q^{\prime}}{2} \right )
\left [
1 + \frac{{q^{\prime}}^2}{2!} + \frac{{q^{\prime}}^2 ({q^{\prime}}^2+2^2)}{4!} \right .\nonumber\\
&\hspace{1cm} \left. + \ldots + \frac{{q^{\prime}}^2 ( {q^{\prime}}^2 + 2^2 ) \ldots ( {q^{\prime}}^2 +(2i-2)^2)}{(2i)!} \right ]
\label{eq:def_alphaeven_discr_sampleddata}\\
&\alpha^{\prime\prime}_{2i+1,fd,d}
=
1 - q^{\prime\prime}\exp\left (- \frac{\pi q^{\prime\prime}}{2} \right )
\left [
1 + \frac{{q^{\prime\prime}}^2 + 1^2}{3!} + \ldots 
\right .\nonumber\\
&\left. 
\hspace{1cm} + \frac{({q^{\prime\prime}}^2 + 1^2)({q^{\prime\prime}}^2 + 3^2)\ldots ({q^{\prime\prime}}^2+(2i-1)^2)}{(2i+1)!} \right ]
\label{eq:def_alphaodd_discr_sampleddata}
\end{align}
in which $q^{\prime(\prime)} \stackrel{\Delta}{=} 2\beta^{\prime(\prime)}/\Delta\tau \gg 1$ is double the normalized stir correlation time. 
Since $\alpha^{\prime(\prime)}_{m,fd,d} < 1$, it follows that imposing the Nyquist criterion further reduces $\lambda^{\prime(\prime)}_m$. 
In particular, from (\ref{eq:def_alphaeven_discr_sampleddata}),
\begin{align}
\alpha^\prime_{0,fd,d} &= 
1 - \exp( -\pi \beta^\prime/\Delta\tau ) < 1 \label{estlambdap0_discr_fd}
\end{align}
i.e., the normalized RMS power $\lambda^\prime_{0,fd,d}$ is negatively biased, unlike $\lambda^\prime_{0,fd,c}\equiv 1$, prompting a renormalization of $g_E(\varpi)$.

\begin{figure}[htb] 
\begin{center}
\begin{tabular}{c}
\vspace{-0.5cm}\\
\hspace{-0.5cm}
\includegraphics[scale=0.65]{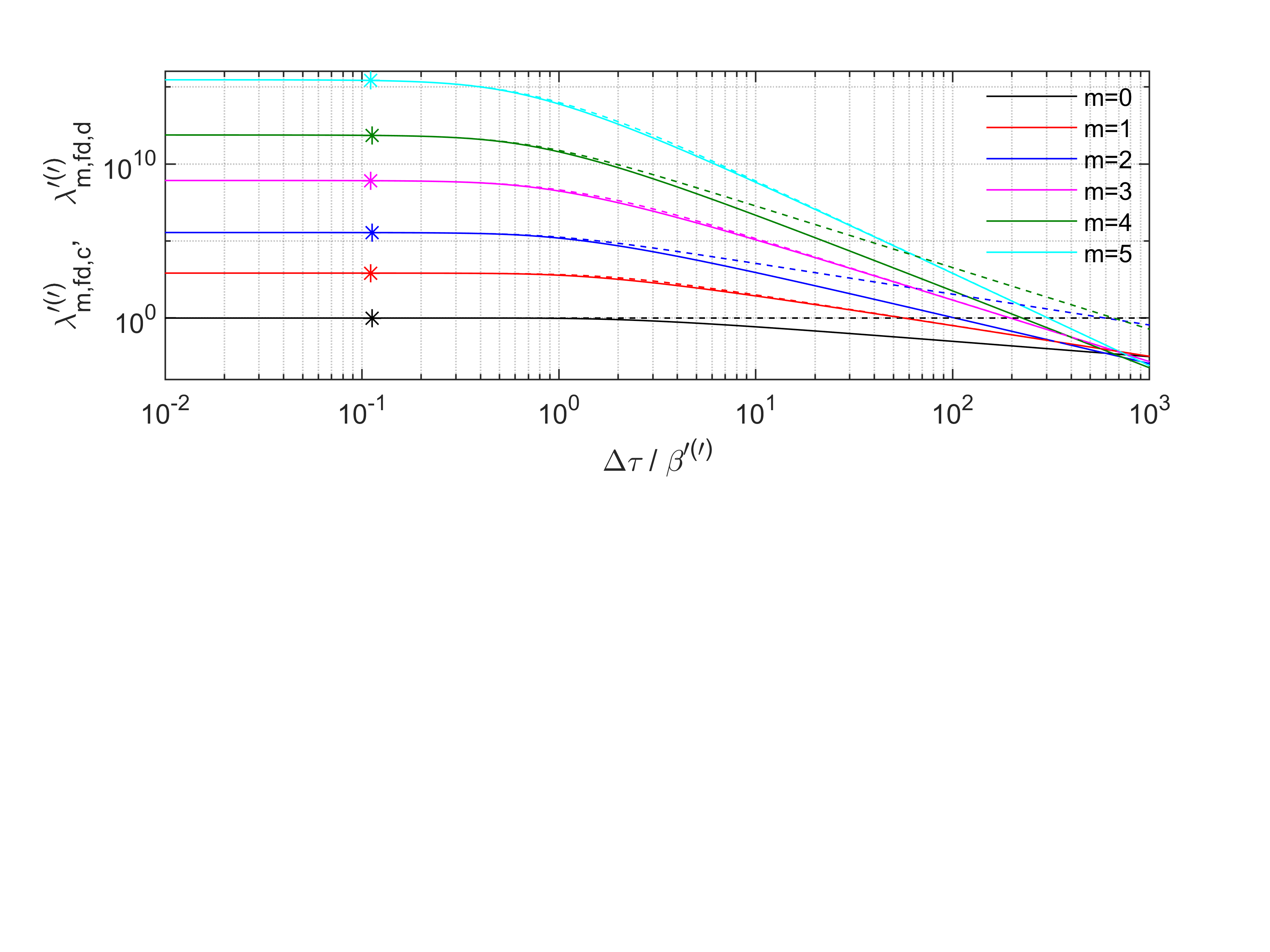}\\ 
\vspace{-4.5cm}\\
\\
(a)\\
\vspace{-0.5cm}\\
\hspace{-0.5cm}
\includegraphics[scale=0.65]{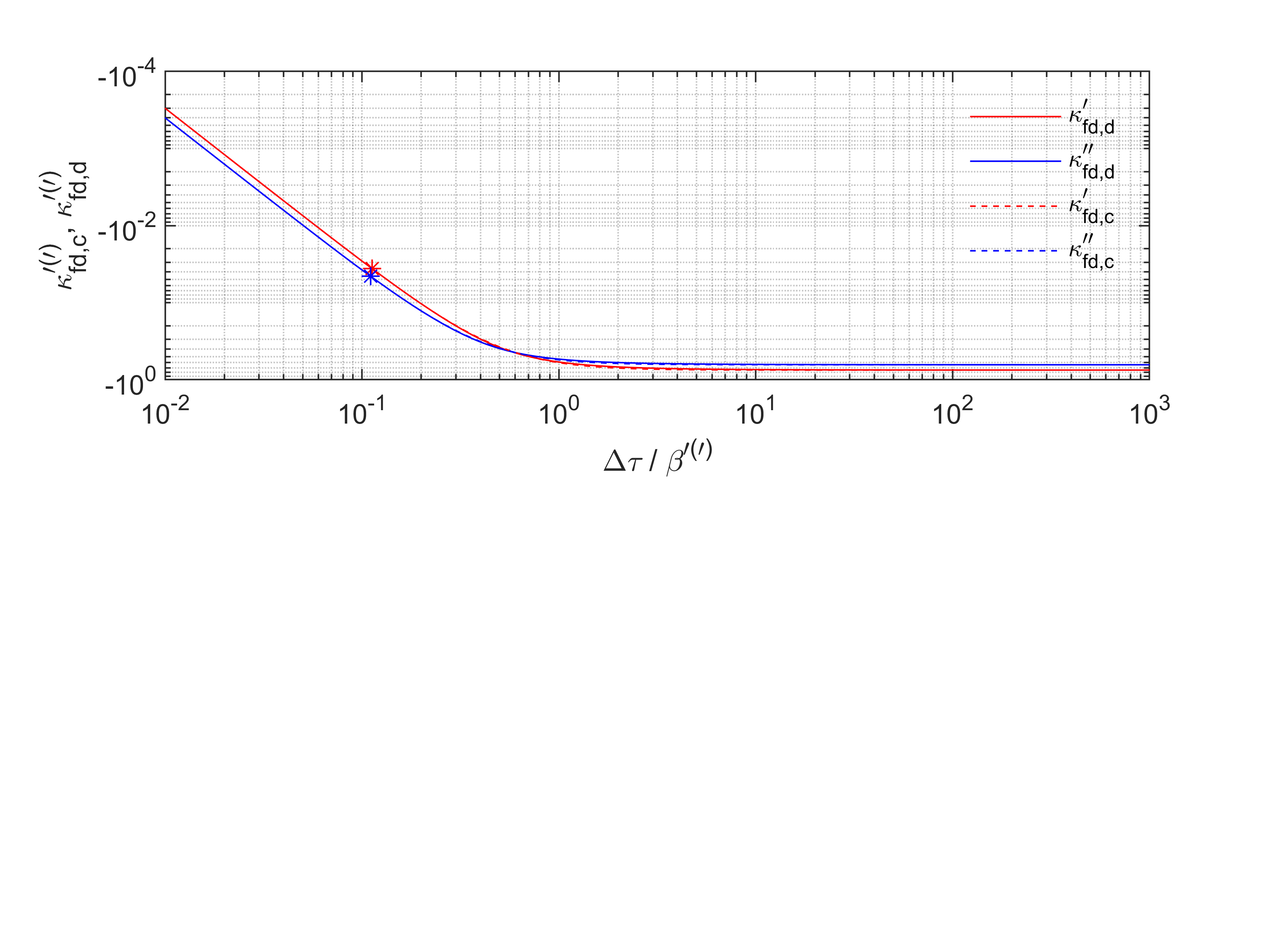}\\ 
\vspace{-4.5cm}\\
\\
(b)\\
\end{tabular}
\end{center}
{
\caption{\label{fig:Fig00_lambdakappa_ifv_Deltatau_FDbias_comp}
\small
{
(a) Spectral moments 
$\lambda^{\prime(\prime)}_{m,fd,c/d}$ 
for $m=0,\ldots 5$, in units (rad/s)$^m$; 
(b) stir parameters $\kappa^{\prime(\prime)}_{fd,c/d}$, based on FD of ideal continuous (dashed; subscript $c$) or actual sampled (solid; subscript $d$) stir sweeps.
Asterisks indicate nominal values of $\Delta\tau/\beta^\prime=0.1128$ 
and $\Delta\tau/\beta^{\prime\prime}=0.1105$ 
at $f=18$ GHz.
}
}
}
\end{figure}

Fig. \ref{fig:Fig00_lambdakappa_ifv_Deltatau_FDbias_comp} compares 
$\lambda^{\prime(\prime)}_{m,fd,c/d}$ and $\kappa^{\prime(\prime)}_{fd,c/d}$ for $0\leq m \leq 5$  as a function of $2/q^{\prime(\prime)}\equiv \Delta\tau/\beta^{\prime(\prime)}$ for FD on analog (i.e., continuous) vs. sampled (i.e., discretized) stir traces.
The moment characteristics $\lambda^{\prime(\prime)}_{m,fd,c/d}(\Delta\tau/\beta^{\prime(\prime)})$ are insensitive when $\Delta\tau/\beta^{\prime(\prime)}\ll 1$, in particular for $\Delta\tau/\beta^{\prime(\prime)} \leq 0.1128$ as in our experiments for $1\leq f\leq 18$ GHz. By contrast, $\kappa^{\prime(\prime)}_{fd,c/d}$ are negative, significantly biased from zero, and sensitive to small changes of $\Delta\tau/\beta^{\prime(\prime)}$ when $\Delta\tau/\beta^{\prime(\prime)}\ll 1$. 
Recall that decreasing $f$ increases $\beta^{\prime(\prime)}$, which lowers the FD bias of $\lambda^{\prime(\prime)}_m$ and $\kappa^{\prime(\prime)}$ for arbitrary $\Delta \tau$. 
Hence, the consideration of FD bias at high $f$ offers a conservative estimate. 

\subsubsection{Super-Nyquist Sampling}
The prevention of aliasing in stir data via low-pass pre-filtering in the spectral stir domain requires prior knowledge of the maximum rate of stir fluctuation \cite{arnamaxratefluct} to decide on the appropriate sampling rate. 
Aliasing has previously been analyzed in the context of late time decay \cite{xu2016}. Although stir aliasing cannot be eradicated after measurement, its influence on the spectral moments and SDF can be gleaned from a adaptation of the previous FD analysis. To this end, extending the integration limits in (\ref{eq:expansion_tau_discr}) to $\pm\varpi_s= \pm 2\pi/{\Delta\tau}$ yields the SDF and moments (with subscript $s$) including the contributions by the aliased power between $\pm \varpi_{\rm max}$ and $\pm \varpi_{\rm s}$.\footnote{Note that extending the spectral range also captures out-of-band stir noise.} 
With 
\begin{align}
\lambda^{\prime(\prime)}_{m,fd,s} 
&= \alpha^{\prime(\prime)}_{m,fd,s} \lambda^{\prime(\prime)}_{m,fd,c} 
\label{eq:def_lambda_discr_sampfreq_supNyquist}
\end{align}
where \cite[eq. (3.895.6)]{grad2007}
\begin{align}
\alpha^\prime_{2i,fd,s}
\stackrel{\Delta}{=} 
1 - \exp\left (- {\pi q^\prime} \right )
\label{eq:def_alpha_discr_sampreq_supNyquist},\hspace{0.2cm}
\alpha^{\prime\prime}_{2i+1,fd,s}
\stackrel{\Delta}{=} 
1 + \exp\left (- {\pi q^{\prime\prime}} \right )
\end{align}
this yields the residual aliased contribution (subscript $a$) as a fraction of the sub-Nyquist contribution to $\lambda^{\prime(\prime)}_{m,fd,c}$, i.e., 
\begin{align}
\lambda^{\prime(\prime)}_{m,fd,a} 
&= \alpha^{\prime(\prime)}_{m,fd,a} \lambda^{\prime(\prime)}_{m,fd,c} 
\label{eq:def_lambda_discr__supNyquist}
\end{align}
where
\begin{align}
&\alpha^\prime_{2i,fd,a}
\stackrel{\Delta}{=} 
\exp\left (- \frac{\pi q^{\prime}}{2} \right )
\left [
1 + \frac{{q^{\prime}}^2}{2!} + \frac{{q^{\prime}}^2 ({q^{\prime}}^2+2^2)}{4!} + \ldots~\right .\nonumber\\
&\hspace{0.3cm}\left. + \frac{{q^{\prime}}^2 ( {q^{\prime}}^2 + 2^2 ) \ldots ( {q^{\prime}}^2 +(2i-2)^2)}{(2i)!} \right ]
-
\exp\left (- {\pi q^{\prime}} \right )
\label{eq:def_alphaeven_discr_sampleddata_supNyquist}\\
&\alpha^{\prime\prime}_{2i+1,fd,a}
\stackrel{\Delta}{=}
q^{\prime\prime}\exp\left (- \frac{\pi q^{\prime\prime}}{2} \right )
\left [
1 + \frac{{q^{\prime\prime}}^2 + 1^2}{3!} + \ldots +
\right .\nonumber\\
&\hspace{-2mm} \left. 
\frac{({q^{\prime\prime}}^2 + 1^2)({q^{\prime\prime}}^2 + 3^2)\ldots ({q^{\prime\prime}}^2+(2i-1)^2)}{(2i+1)!} \right ]
+
\exp\left (- {\pi q^{\prime\prime}} \right )
.
\label{eq:def_alphaodd_discr_sampleddata_supNyquist}
\end{align}
For sampled data, $\lambda^{\prime(\prime)}_{m,fd,a} = (\alpha^{\prime(\prime)}_{m,fd,a}/\alpha^{\prime(\prime)}_{m,fd,d})\lambda^{\prime(\prime)}_{m,fd,d} $.

\begin{figure}[htb] 
\begin{center}
\begin{tabular}{c}
\vspace{-0.5cm}\\
\hspace{-0.7cm}
\includegraphics[scale=0.68]{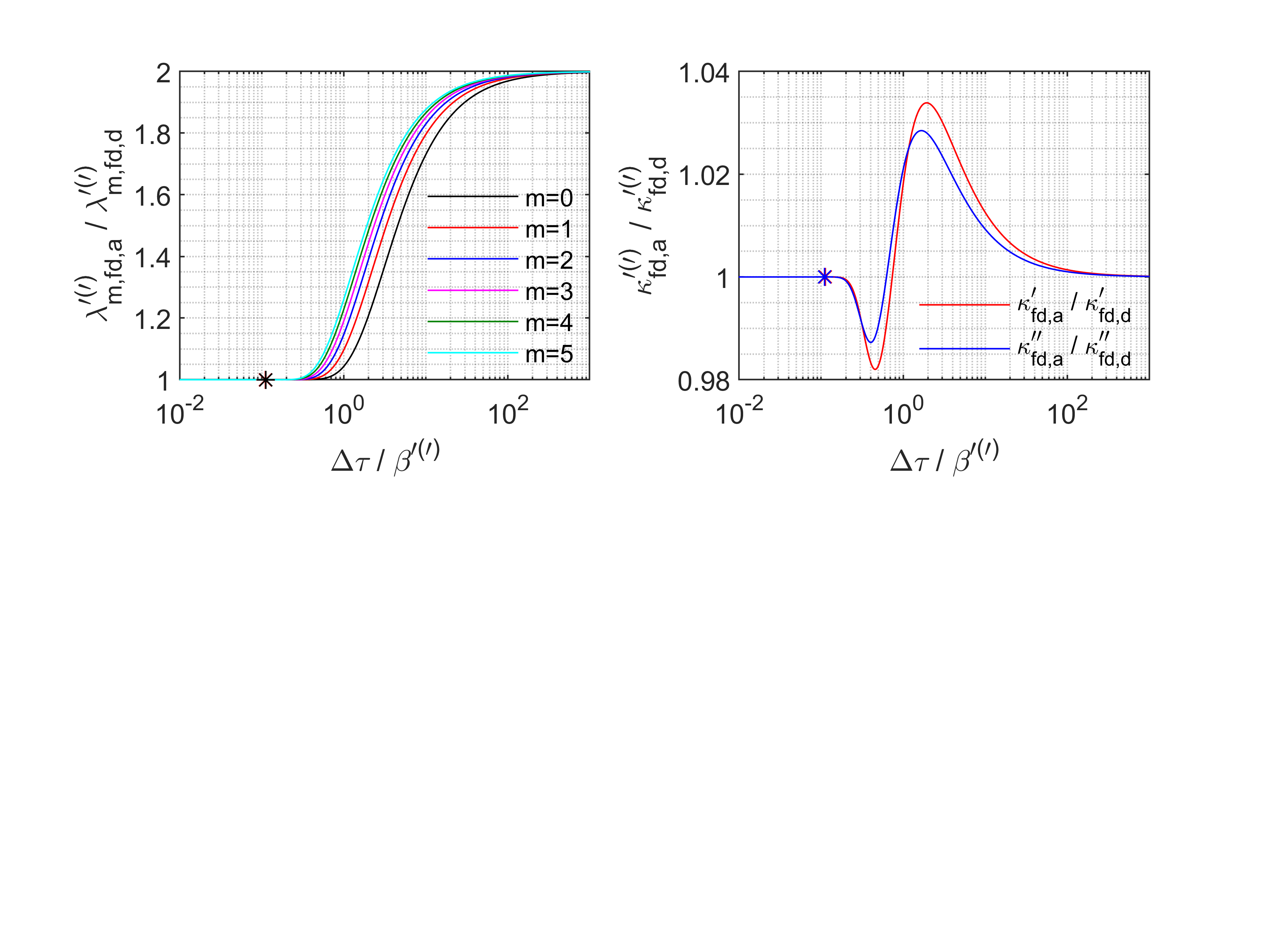}\\ 
\vspace{-4.4cm}\\
\\
(a)~~~~~~~~~~~~~~~~~~~~~~~~~~~~~~(b)
\end{tabular}
\end{center}
{
\caption{\label{fig:Fig00b_lambdakappa_ifv_Deltatau_FDbias_subsuperNyquist}
\small
{
Ratios of aliased to non-aliased spectral parameters: (a) relative $\lambda^{\prime(\prime)}_{m,fd,a}$ for $m=0,\ldots,5$; (b) relative $\kappa^{\prime(\prime)}_{fd,a}$, as a function of $\Delta\tau/\beta^{\prime(\prime)}$.
Asterisks indicate nominal values of $\Delta\tau/\beta^\prime=0.1128$ 
and $\Delta\tau/\beta^{\prime\prime}=0.1105$ 
at $f=18$ GHz.
}
}
}
\end{figure}

Fig. \ref{fig:Fig00b_lambdakappa_ifv_Deltatau_FDbias_subsuperNyquist} shows $\lambda^{\prime(\prime)}_{m,fd,a} / \lambda^{\prime(\prime)}_{m,fd,d}$ and $\kappa^{\prime(\prime)}_{fd,a} / \kappa^{\prime(\prime)}_{fd,d}$ for $0 \leq m \leq 5$ as a function of $\Delta\tau/\beta^{\prime(\prime)}$ at $f=18$ GHz. 
The results indicate a finite transition region, approximately $0.5 < \Delta\tau/\beta^{\prime(\prime)} < 500$. Higher-order moments exhibit an earlier onset of increasing $\lambda^{\prime(\prime)}_{m,fd,a} / \lambda^{\prime(\prime)}_{m,fd,d}$. 
The deviation of $\kappa^{\prime(\prime)}_{fd,a} / \kappa^{\prime(\prime)}_{fd,d}$ from unity is less than $4\%$ across the transition region, and less than $0.1\%$ in our measurements.
Since stir spectra are in general flatter for higher $f$ (larger stir bandwidth), the fraction of aliased power for lower $f$ is expected to be smaller, although the stir LF-to-HF transition of stir power density is then also less abrupt. 

In summary, the estimated effect of stir aliasing on the FD bias of $\kappa^{\prime(\prime)}$ is negligible for $f\leq18$ GHz, being just a few percent or less at the highest frequency, for arbitrary $\Delta\tau/\beta^{\prime(\prime)}$. 

The foregoing analysis was performed in the stir-spectral DTFT domain for continuous $\varpi$. It is easily extended to DFTs for discrete $\varpi_k$ by replacing integrations with sums over $\varpi_k = k \Delta \varpi = 2 \pi k / (N_s\Delta \tau)$,
where $k=0,1,\ldots,N_s - 1$.

\subsection{Estimation Based on Complex Autocorrelation\label{subsec:FDbiasCov}}
In sec. \ref{sec:contstirsweep}, independent estimates of $\beta^{\prime(\prime)}$ were sought for estimating FD bias. These can be provided by the ACV (or ACF) method \cite{mill1972}. This alternative technique for estimating any $\lambda_m$ still operates within the temporal stir domain, but uses the $m$th-order FD of the complex ACF instead, i.e.,
\begin{align}
\rho_E(\tau) \equiv a_{\rho_E}(\tau) \exp[\rmj \phi_{\rho_E}(\tau)]
\end{align} 
for complex $E(\tau)$,
where $a_{\rho_E}(\tau) =  [\rho^{\prime^2}_E(\tau)+\rho^{\prime\prime^2}_E(\tau)]^{1/2}$ and $\phi_{\rho_E}(\tau) =  \tan^{-1}(\rho^{\prime\prime}_E(\tau)/\rho^{\prime}_E(\tau))$ are both to be evaluated at zero lag,
or alternatively using the ACV function $\gamma_E(\tau)$. Thus,
\begin{align}
{\lambda_m} 
= (-\rmj)^m {\rho^{(m)}_E}(0) \sigma^2_E 
= 
(-\rmj)^m {{\gamma^{(m)}_E}(0)}
.
\label{eq:covmethod}
\end{align}
Extending the results for $m\leq 2$ obtainable from \cite{mill1972} 
to the first six moments that govern $\kappa^{\prime(\prime)}$ yields
\begin{align}
\lambda^\prime_0 &= a_{\rho_E}(0) = 1\label{eq:lambda0_complexACFmethod}\\
\lambda^{\prime\prime}_1 &= \dot{\phi}_{\rho_E}(0)
\label{eq:lambda1_complexACFmethod}\\
\lambda^\prime_2 &= - \ddot{a}_{\rho_E}(0) + (\lambda^{\prime\prime}_1)^2
\label{eq:lambda2_complexACFmethod}\\
\lambda^{\prime\prime}_3 &= - \dddot{\phi}_{\rho_E}(0) - 2 (\lambda^{\prime\prime}_1)^3 + 3 \lambda^{\prime\prime}_1 \lambda^\prime_2
\label{eq:lambda3_complexACFmethod}\\
\lambda^{\prime}_4 &= {a}^{(4)}_{\rho_E}(0) + 3 (\lambda^{\prime\prime}_1)^4 - 6 (\lambda^{\prime\prime}_1)^2\lambda^\prime_2 + 4 \lambda^{\prime\prime}_1 \lambda^{\prime\prime}_3
\label{eq:lambda4_complexACFmethod}\\
\lambda^{\prime\prime}_5 &= 
{\phi}^{(5)}_{\rho_E}(0) 
- 24 (\lambda^{\prime\prime}_1)^5 
+ 60 (\lambda^{\prime\prime}_1)^3\lambda^\prime_2 
- 20 (\lambda^{\prime\prime}_1)^2\lambda^{\prime\prime}_3 
\nonumber\\
&~~~~+ 5 \lambda^{\prime\prime}_1 \lambda^{\prime}_4 - 30 \lambda^{\prime\prime}_1 (\lambda^{\prime}_2)^2 
+ 10 \lambda^{\prime}_2
\lambda^{\prime\prime}_3 
.
\label{eq:lambda5_complexACFmethod}
\end{align}
Since the complex ACF exhibits Hermitean symmetry, i.e., 
\begin{align}
a_{\rho_E}(\tau) = a_{\rho_E}(-\tau),~~~\phi_{\rho_E}(\tau)=-\phi_{\rho_E}(-\tau)
\label{eq:complexACF_Hermitean}
\end{align} 
this can be exploited to halve the number of lags $\tau_n$ required for
estimating $\lambda^{\prime(\prime)}_{m}$. Specifically, for FD  of sampled data
\begin{align}
\dot{\phi}_{\rho_E}(0) &\simeq \frac{2 \left [ {\phi}_{\rho_E}(\Delta\tau) - {\phi}_{\rho_E}(0) \right ]}{2\Delta\tau} \equiv \frac{ {\phi}_{\rho_E}(\Delta\tau) }{\Delta\tau} \label{eq:phi0_complexACFmethod}
\\
\ddot{a}_{\rho_E}(0) &\simeq \frac{2\left [ {a}_{\rho_E}(\Delta\tau) - {a}_{\rho_E}(0) \right ]}{(\Delta\tau)^2} \equiv \frac{2\left [ {a}_{\rho_E}(\Delta\tau) - 1 \right ]}{(\Delta\tau)^2}\\
\dddot{\phi}_{\rho_E}(0) &\simeq \frac{2\left [ {\phi}_{\rho_E}(3\Delta\tau) - 3 {\phi}_{\rho_E}(\Delta\tau) \right ]}{(2\Delta\tau)^3}\\
{a}^{(4)}_{\rho_E}(0) &\simeq \frac{2\left [ {a}_{\rho_E}(2\Delta\tau) - 4 {a}_{\rho_E}(\Delta\tau) + 3 {a}_{\rho_E}(0) \right ]}{(\Delta\tau)^4}\\
{\phi}^{(5)}_{\rho_E}(0) &\simeq \frac{2\left [ {\phi}_{\rho_E}(5\Delta\tau) - 5 {\phi}_{\rho_E}(3\Delta\tau) + 10 {\phi}_{\rho_E}(\Delta\tau) \right ]}{(2\Delta\tau)^5}
\label{eq:phi5_complexACFmethod}
\end{align}
etc. 
Because of the high oversampling rate offered by a VNA, multi-point ACF estimation schemes \cite{pass1983} for complementing (\ref{eq:phi0_complexACFmethod})--(\ref{eq:phi5_complexACFmethod}) offer no further advantage, even for large stir bandwidths.
Since the order of differentiation for $\rho_E(0)$ in (\ref{eq:covmethod}) equals the moment order $m$, an FD bias compensation similar to that for stir sweeps in sec. \ref{subsec:FDbiasSweep} can be applied.

\subsection{Estimation Based on Spectral Density Function}
Naturally, spectral moments can also be estimated in the spectral stir domain, from their definition (\ref{eq:def_lambdam_derivatives_temp}) \cite{mill1970b}.
Estimating $\cE(\varpi) \cE^*(\varpi)$ based on a DFT of $E(\tau)$, applying the forward transformation in (\ref{eq:periodogram_discr}), with $N_s \Delta \varpi/2 = \pi/\Delta \tau$, and following (\ref{eq:def_lambdaeven_discr}), this yields 
\begin{align}
\lambda_m \simeq \sigma^2_E
\sum^{{N_s}/{2}-1}_{k=0} [{\varpi_k}  \sinc (\pi k / N_s)]^m \hat{g}_E(\varpi_k) \Delta\varpi
.
\label{eq:def_lambdai_discr_DFT}
\end{align}

However, it is well known that this spectral method is in general ill-conditioned.
An FFT performed on a finite record of sampled stir data or correlation series causes its sample SDF to cover only a finite stir band $N_s\Delta\varpi$, resulting in a biased estimate of the sought continuous unrestricted SDF.
Sample spectrum estimates are in general neither unbiased (except for ideal white noise) nor consistent.
On the other hand, by the Nyquist theorem, a periodogram-based SDF estimate is aliased when it is not bandlimited. 
For the moments, a bias in the estimated SDF causes in turn a bias of $\lambda_m$ when derived from (\ref{eq:def_lambdai_discr_DFT}),
{\em a fortiori} for large spectral bandwidths
approaching $\pi/\Delta\tau$ \cite{zrni1979}, 
\cite{meln2004}.
Especially higher-order $\lambda_m$ increasingly depend on the SDF tail and are, hence, increasingly affected by aliasing or truncation.
Furthermore, the standard uncertainty of $\Lambda_m$  rapidly increases with $m$ and is larger than in temporal estimation \cite{zrni1979}. 
Finally, in spectral estimation, $\tilde{\lambda}_m$ is merely asymptotically unbiased, i.e., for $T\rightarrow +\infty$ in a single stir sweep, which is not applicable to rotational periodic stirring.

In general, widening the IFBW $B$ and stir bandwidth $\sqrt{\lambda^\prime_2/2}$ (both in units rad/s) results increasingly in spectral nonuniformity for the bias of the ASDF, leading to the peaks and troughs in the ASDF becoming under- and overestimated, respectively. 
The bias error of the ASDF, $b_{g_E}$, is strongly nonuniform when $B \ll \sqrt{\lambda^\prime_2/2}$: for the $[0/1]$ model \cite{bend1986}
\begin{align}
b_{g_E}(\varpi) \simeq \frac{B^2 \ddot{g}_E(\varpi)}{96\pi^2} \propto \frac{B^2 }{ (\lambda^\prime_2)^{3/2}} \exp \left ( - \sqrt{{2}/{\lambda^\prime_2}} \varpi \right )
\label{eq:bis_gE_1storder}
\end{align}
i.e., decreasing exponentially.
This strong nonuniformity also leads to unequal bias of estimated $\lambda_m$ for different $m$. 
Note, however, that the relative bias $\varepsilon_b ={b_{g_E}(\varpi)}/{g_E(\varpi)} \propto B^2 / \lambda^\prime_2$ is independent of $\varpi$ in this model.

\section{Bias Caused by Noise, Interference, or Understirring}

\subsection{Stir Noise\label{sec:stirnoisekurt}}
The FD bias analyzed in sec. \ref{sec:FDbias} is a measurement artefact and by-product of sampling by the VNA. 
Physical imperfections to ideal randomization of the continuous field itself inside the MSRC, prior to sampling, are now investigated for their contribution to spectral moments. Unlike for FD bias, physical imperfections are more difficult to compensate in spectral moments and kurtosis, because they require more detailed information about field contributions, as shown below.
For brevity, only the real ASDF $g^\prime_E(\varpi)$ is analyzed here.
 
\subsubsection{Continuous Ideal Stirred Field and Baseband Stir Noise\label{subsubsec:FieldPlusNoise}}
\begin{figure}[htb] 
\begin{center}
\begin{tabular}{c}
\hspace{-0.5cm}
\includegraphics[scale=0.63]{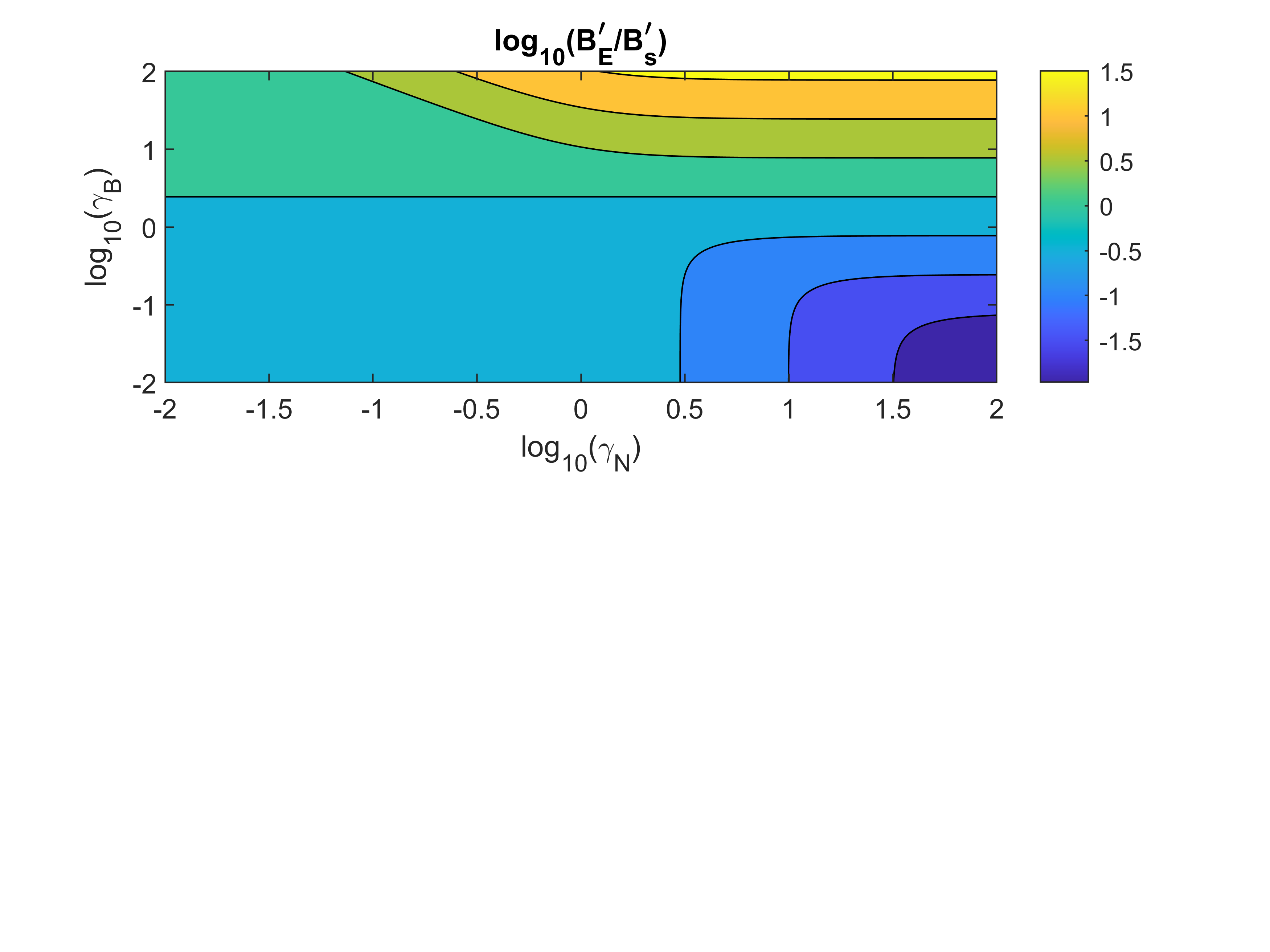}\\
\vspace{-4.1cm}\\ 
(a)\\
\\
\vspace{-0.7cm}\\
\hspace{-0.5cm}
\includegraphics[scale=0.63]{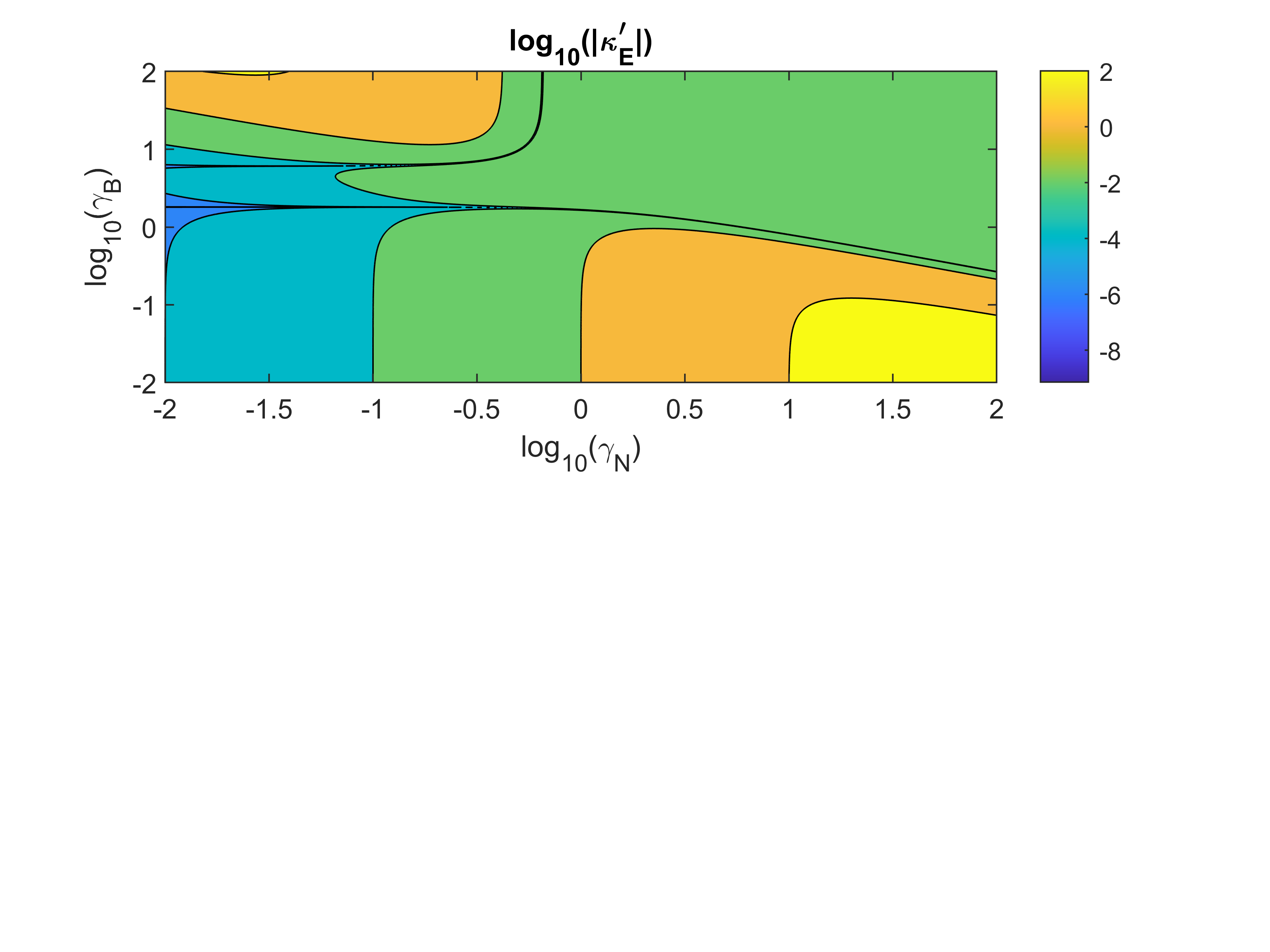}\\
\vspace{-4.1cm}\\ 
(b)\\
\\
\vspace{-0.7cm}\\
\hspace{-0.7cm}
\includegraphics[scale=0.67]{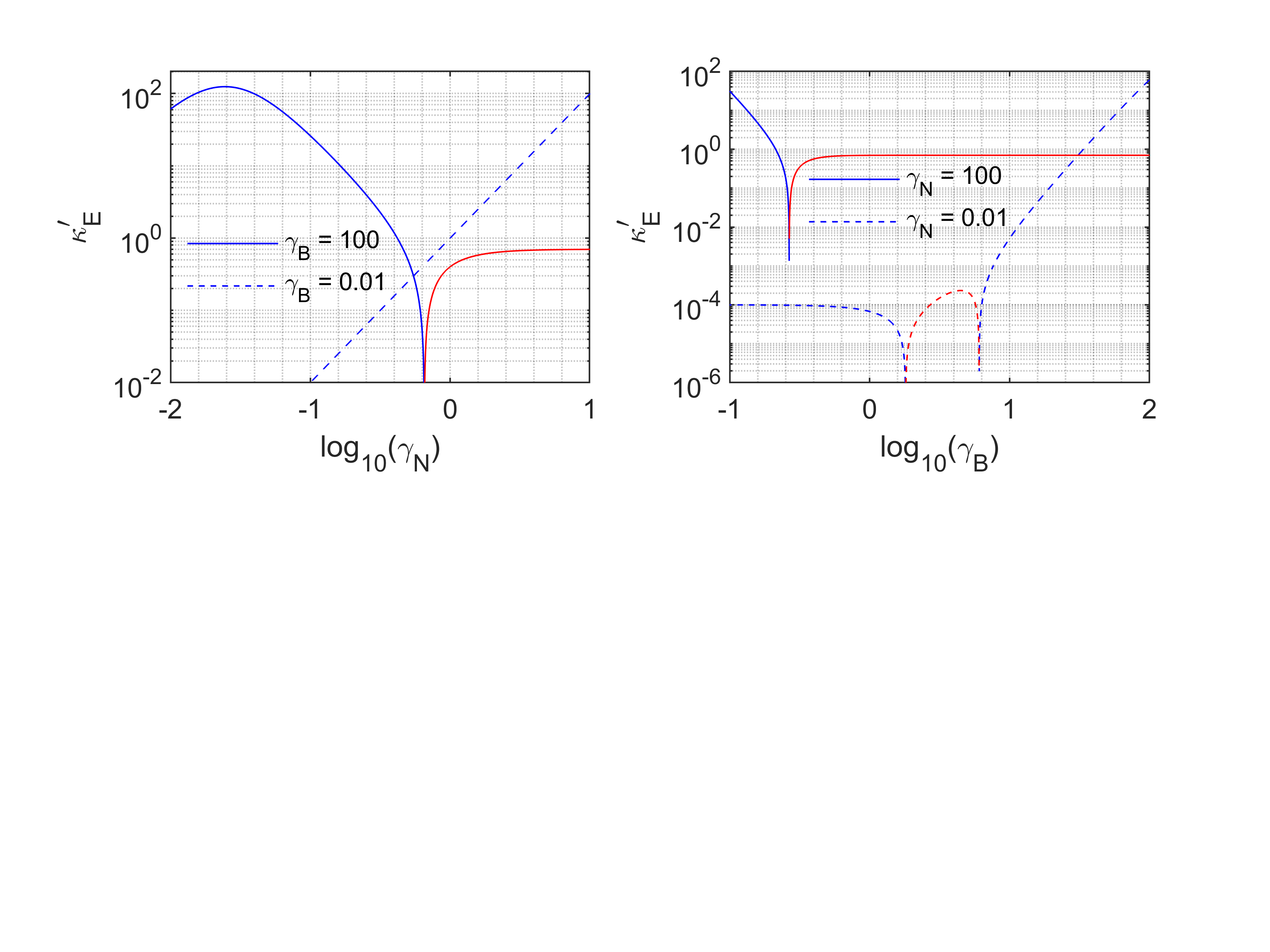}\\
\vspace{-4cm}\\
~~(c)~~~~~~~~~~~~~~~~~~~~~~~~~~~~~~~~~~(d)
\\
\end{tabular}
\end{center}
{
\caption{\label{fig:kappap_ifv_SNR}
\small
{
(a)
${\rm log}_{10}(\cB^\prime_E/\cB^\prime_s)$
and
(b)
${\rm log}_{10}(|\kappa^\prime_E|)$
as functions of NSRs for RMS levels and bandwidths, ${\rm log}_{10}(\gamma_N)$ and ${\rm log}_{10}(\gamma_B)$;
(c) $\kappa^\prime_E(\gamma_N)$ and (d) $\kappa^\prime_E(\gamma_B)$ for selected $\gamma_B$ and $\gamma_N$, respectively, indicating positive (blue) and negative (red) values.
}
}
}
\end{figure}

Consider a random noise field, $N(\tau)$, independent of but additive to the stirred field $E_s(\tau)$ \cite[eq. (81)]{arnaACFSDF_pt1}, i.e.,
\begin{align}
E(\tau) = E_s(\tau) + N(\tau)
.
\label{eq:EisE0plusN}
\end{align}
Such $N(\tau)$ may have an electrical, electromechanical, thermoelectric, or other origin. It is modelled here as bandlimited baseband white noise with $g^\prime_N(\varpi)=1/\cB^\prime_N$ for $0 \leq \varpi \leq \cB^\prime_N$. 
On account of the additivity of ACV functions and Bochner's theorem \cite{wong1985}, the non-normalized ASDF of $E(\tau)$ is
\begin{align}
\sigma^2_E g^\prime_E(\varpi) = \sigma^2_{E_s} g^\prime_{E_s}(\varpi) + \sigma^2_N g^\prime_N(\varpi)
\label{eq:SDF_EisE0plusN}
\end{align}
and $\sigma^2_E = \sigma^2_{E_s} + \sigma^2_N$.
Denote $\cB^\prime_s$ as the stir bandwidth of $E_s(\tau)$, where $\cB^\prime_s \equiv 1/\beta^\prime_s = (\lambda^\prime_{2,E_s}/2)^{1/2}$ with
 $g^\prime_{E_s}(\varpi) = \exp(-\varpi / \cB^\prime_s) / \cB^\prime_s$ for ideal stirred $E_s(\tau)$ and $0 < \varpi < + \infty$.
With (\ref{eq:specmom_ideal1storderPade}), the spectral moments of $E(\tau)$ are then
\begin{align}
\lambda^{\prime}_{m,E} 
&= {m!} {(\cB^\prime_s)^{m}} \frac{\sigma^2_{E_s}}{\sigma^{2}_E} + \frac{(\cB^\prime_N)^m}{m+1} \frac{\sigma^2_{N}}{\sigma^{2}_E}  \label{eq:lambdan_E0plusN_temp}\\
&=  {m!} (\cB^\prime_s)^{m}\frac{1 + \gamma^{2}_N \gamma^{m}_B / (m+1)!}{1+\gamma^{2}_N}
\label{eq:lambdan_E0plusN}
\end{align}
where 
\begin{align}
\gamma_B \stackrel{\Delta}{=} \cB^\prime_N / \cB^\prime_s,\hspace{0.3cm}
\gamma_N \stackrel{\Delta}{=} \sigma_N / \sigma_{E_s}
\label{eq:def_gammaB_gammaN_EplusN}
\end{align}
denote the noise-to-stir ratios (NSRs) of the bandwidths and RMS amplitudes, respectively, between $N(\tau)$ and $E_s(\tau)$. 
The overall bandwidth $\cB^\prime_E = ({\lambda^\prime_{2,E}}/{2})^{1/2} $ and kurtosis $\kappa^\prime_E = {\lambda^\prime_{4,E}} / [{6(\lambda^\prime_{2,E})^2}] - 1 $ for $E(\tau)$ then follow as 
\begin{align}
\cB^\prime_E &= 
\cB^\prime_s \sqrt{\frac{1+\gamma^2_N \gamma^2_B/6}{1+\gamma^2_N}}\label{eq:BWp_stirplusnoise}
\end{align}
\begin{align}
\kappa^\prime_E 
&= \frac{\gamma^2_N}{\left ( 1 + {\gamma^2_N \gamma^2_B}/{6} \right )^2}
\left ( 
1 - \frac{\gamma^2_B}{3} + \frac{\gamma^4_B}{120} - \frac{7\gamma^2_N \gamma^4_B}{360} \right )
\label{eq:kappap_E0plusN}
.
\end{align}
Fig. \ref{fig:kappap_ifv_SNR}(b) indicates that $\kappa^\prime_E(\gamma_N,\gamma_B)$
exhibits loci of matching pairs $(\gamma_N,\gamma_B)$ where $\kappa^\prime_E=0$, which are related via
\begin{align}
1 - 40\gamma^{-2}_B + 120\gamma^{-4}_B = 7 \gamma^{2}_N /3
.
\label{kappap_E0plusN_zeroes}
\end{align}

If the RMS noise level overwhelms the RMS magnitude of the stirred field ($\gamma_N \gg 1$), then 
\begin{align}
\kappa^\prime_E \simeq -7/10 + [3/10 - (18/5) \gamma^{-2}_B + 36 \gamma^{-4}_B ] \gamma^{-2}_N 
.
\label{kappap_E0plusN_smallgammaN}
\end{align} 
Conversely, for vanishing RMS noise levels ($\gamma_N \rightarrow 0$) and bandwidths ($\gamma_B \rightarrow 0$),
$\kappa^\prime_E$ represents asymptotically the NSR of power levels, viz.,
\begin{align}
\kappa^\prime_E \simeq \left ( 1 - \gamma^{2}_B/3 + \gamma^{4}_B / 120 \right ) \gamma^{2}_N \rightarrow \gamma^{2}_N 
.
\label{eq:E0plusN_inverseSNR}
\end{align} 
For typical VNA measurements inside efficiently stirred MSRCs ($\cB^\prime_N\gg \cB^\prime_s$, $\sigma_N \ll \sigma_{E_s}$), we have $\gamma_N \ll 1$ and\footnote{While the bandwidth is manually selected on the analyzer, the bandwidth of a mechanical stir process is typically much lower than that of $N_e(\tau)$.} $\gamma_B \gg 1$. 
In this regime, Fig. \ref{fig:kappap_ifv_SNR}(c) indicates that $\kappa^\prime_E(\gamma_N)$ is predominantly positive but rapidly decreasing with increasing $\gamma_N$. 
Conversely, $\kappa^\prime_E(\gamma_B)$ also increases by several orders of magnitude to high positive values, possibly with one or two zero crossings, as is apparent from Fig. \ref{fig:kappap_ifv_SNR}(d). Thus, typically, $\kappa^\prime_E$ is positive but highly sensitive to variations of $\gamma_N$ and $\gamma_B$. 

By extension, for alias-free sampled $E_s(\tau_n)$, the first term in (\ref{eq:lambdan_E0plusN_temp}) generalizes to
\begin{align}
\left [ {m!}{(\cB^\prime_s)^{m}} - \sum^m_{i=0}\frac{m! ~\varpi^{m-i}_{\rm max} (\cB^\prime_s)^{i} \exp(- \varpi_{\rm max}/ \cB^\prime_s )}{(m-i)! }  \right ] \frac{\sigma^2_{E_s}}{\sigma^{2}_E} 
.
\label{eq:lambdan_E0plusN_samp}
\end{align}
If the stir noise is not baseband (i.e., not coupled to stir DC; cf. sec. \ref{sec:understirringkurt}) but restricted to some finite band $[\cB^\prime_{N,\ell},\cB^\prime_{N,u}]$, then $(\cB^\prime_N)^m$ in (\ref{eq:lambdan_E0plusN_temp}) is to be  replaced with $[(\cB^{\prime}_{N,u})^{m+1} - (\cB^{\prime}_{N,\ell})^{m+1}]/(\cB^\prime_{N,u}-\cB^\prime_{N,\ell})$, with corresponding changes to (\ref{eq:def_gammaB_gammaN_EplusN})--(\ref{eq:kappap_E0plusN}).

\subsubsection{Stir Noise vs. Electrical Noise\label{sec:stirnoise}}
Stir noise $N_s(\tau)$ represents contributing random fluctuations that vary at a rate that is much faster than that for the stirred field process $E_s(\tau)$ itself. Such noise may have different physical origins, also including quantization noise. 
Typically, $N_s(\tau)$ varies more slowly than  electrical noise $N_e(t)$ that is already present in the absence of stirring. 
Observations of typical VNA stir sweeps at high SNR level ratios indicate that $N_s(t)$ resembles shot noise, i.e., appearing as Poissonian, approximately white and additive, $N(t)=N_s(t)+N_e(t)$.

Another case arises when signal and noise intensities are of comparable RMS magnitude, i.e., $\overline{E^2_s(\tau)} \sim \overline{N^2(\tau)}$, {\it a fortiori} when $\overline{E^2_s}$ is small and $N(t)$ is the thermal noise field \cite{marv2022}. Then, stirring of this thermal noise must be taken into account (stirred $N_e(t)$). With all interior surfaces of the MSRC then radiating at nonzero temperature, $g_N(\varpi)$ then exhibits a significant unstirred component.

Pure nonelectrical contributions may be separated from electrical noise using static or (ideal) mode tuning, provided that the tuner is sufficiently rigid in order to produce either a mechanically truly static state,  
or to move at a rate that is much slower than the rates of the electrical noise and cavity field relaxation (decay) \cite{arnamaxratefluct}.

\subsection{Harmonic EMI and Understirred Fields \label{sec:understirringkurt}}
To evaluate the contribution of permanent periodic interfering fields on the spectral moments, consider $E(\tau)$ to be the superposition of a perfectly stirred $E_s(\tau)$ and a discrete set of deterministic harmonic EMI contributions $E_h(\tau)$
\begin{align}
E(\tau) = E_s(\tau) + \sum_{h} E_h(\tau).
\label{eq:E0constributions}
\end{align} 
Whilst $g_{E_s}(\varpi) = \exp(-\varpi / \cB^\prime_s ) / \cB^\prime_s$ remains continuous,
each spectral EMI component $\cE_h(\varpi) = \cE_{h,0} \exp(\rmj \phi_h) \delta(\varpi-\varpi_h)$ originating from the $\omega$-domain adds a discrete contribution $g_{E_h}(\varpi) = \delta(\varpi-\varpi_h)$ to the SDF of $E_s(\tau)$. 
Proceeding as in sec. \ref{sec:stirnoisekurt}, with $\sigma^2_E = \sigma^2_{E_s} + \sum_h \sigma^2_{E_h}$, the ASDF of $E(\tau)$ and its moments 
follow from (\ref{eq:EisE0plusN}), (\ref{eq:SDF_EisE0plusN}) and (\ref{eq:E0constributions}) as
\begin{align}
\sigma^2_{E} g^\prime_{E}(\varpi) 
&= \sigma^2_{E_s} g^\prime_{E_s}(\varpi) + \sum_{h} |\cE_{h,0}|^2 g^\prime_{E_h}(\varpi) \\
\lambda^{\prime}_{m,E} &= {m!} (\cB^\prime_s)^m \frac{\sigma^2_{E_s}}{\sigma^{2}_{E}} + \sum_{h} \varpi^m_h \frac{|\cE_{h,0}|^2 }{\sigma^{2}_{E}} 
.
\label{eq:E0plusEh_specmomgen}
\end{align}
In particular, if EMI is confined to a single field harmonic then
\begin{align}
\lambda^{\prime}_{m,E} &= {m!} (\cB^\prime_s)^{m} \frac{1 + \gamma^{2}_{E_h} \gamma^{m}_{\varpi_h} / m!}{1+\gamma^{2}_{E_h}}
\label{eq:E0plusEh_singleline}
\end{align}
\begin{align}
\cB^\prime_E &= 
\cB^\prime_s \sqrt{\frac{1+\gamma^2_{E_h} \gamma^2_{\varpi_h}/2}{1+\gamma^2_{E_h}}}\label{eq:BWp_stirplusnoise}
\end{align}
\begin{align}
\kappa^\prime_E &= \frac{\gamma^2_{E_h}}{\left ( 1 + \gamma^2_{E_h} \gamma^2_{\varpi_h}/{2} \right )^2}
\left ( 
1 - \gamma^2_{\varpi_h} + \frac{\gamma^4_{\varpi_h}}{24} - \frac{5\gamma^2_{E_h} \gamma^4_{\varpi_h}}{24} \right )
\label{eq:kappap_E0plusEh}
\end{align}
in which the interference-to-stir ratios (ISRs) for frequencies and RMS amplitudes are defined by
\begin{align}
\gamma_{\varpi_h} \stackrel{\Delta}{=} {\varpi_h} / \cB^\prime_s ,~~~
\gamma_{E_h} \stackrel{\Delta}{=} |\cE_{h,0}| / \sigma_{E_s}
.
\label{eq:ISR_BWandAmp}
\end{align}
If the EMI is random or intermittent, then the amplitude ISR can be redefined as
$\gamma_{E_h} \stackrel{\Delta}{=}  \sigma_{E_h} / \sigma_{E_s}
$ with $\sigma_{E_h} = \langle |\cE_{h}|^2 \rangle^{1/2}$.

Since (\ref{eq:kappap_E0plusEh}) is formally similar to (\ref{eq:kappap_E0plusN}), the plot of $\kappa^\prime_E(\gamma_{{E_h}},\gamma_{\varpi_h})$ \cite{arnaATRASC2024} is similar to $\kappa^\prime_E(\gamma_N,\gamma_B)$ in Fig. \ref{fig:kappap_ifv_SNR}(b), now with its zero crossings occurring for
\begin{align}
1 - 24\gamma^{-2}_{\varpi_h} + 24\gamma^{-4}_{\varpi_h}  = {5} \gamma^{2}_{E_h} 
.
\label{kappap_E0plusEh_zeroes}
\end{align}

Some special cases of (\ref{eq:kappap_E0plusEh}) can now be considered.
Firstly, an unstirred field of stir-spectral magnitude $\cE_{u,0}$ represents the stir DC limit of a stir `harmonic', i.e., $\varpi_h =0\stackrel{\Delta}{=}\varpi_u$.
Hence $\cE_u(\varpi) = \cE_{u,0} \exp(\rmj \phi_u) \delta(\varpi)$, in which $\cE_{u,0}$ and/or $\phi_u$ are (quasi-)stationary or deterministic relative to $\cE_s(\varpi)$, by the same token as for $\sigma^2_{E_h}$ before. After normalization by $|\cE_{u,0}|$ or $\langle |\cE_{u,0}|^2 \rangle^{1/2}$, its SDF contribution is $g_{E_u}(\varpi) = \delta(\varpi)$.
From (\ref{eq:kappap_E0plusEh}) with $\gamma_{E_u}\gamma_{\varpi_u} \gg 1$, regardless of the magnitude of $\gamma_{E_u}$, 
\begin{align}
\kappa^\prime_{E}\simeq \left ( 1 - \gamma^{2}_{\varpi_u} + \gamma^{4}_{\varpi_u}/24\right ) \gamma^{2}_{E_u} \rightarrow \gamma^{2}_{E_u}
\label{eq:kappa_E0plusEu}
\end{align}
in which $\gamma^{2}_{E_u}$ is the familiar Ricean $K$-factor.
Naturally, (\ref{eq:kappa_E0plusEu}) parallels (\ref{eq:E0plusN_inverseSNR}) because baseband stir noise with a vanishingly small bandwidth ($\gamma_B \rightarrow 0$) is equivalent to an unstirred or quasi-statically stirred (i.e., weakly stir modulated) field contribution ($\gamma_{\varpi_u} \rightarrow 0$) to  $E(\tau)$.
In this case, from (\ref{eq:kappa_E0plusEu}), $\kappa^\prime_{E}$ increases for an increasing level $\cE_{u,0}$ or $\sigma_{E_{u}}$. This 
result agrees with one interpretation of kurtosis as representing `peakedness' in a density function. 
Thus, in the absence of stir noise, understirring gives rise to a positive contribution to $\kappa^\prime_{E}(\gamma_{E_u})$. 

As a second case, for a dominant non-DC quasi-harmonic contribution at a high stir frequency ($\gamma_{E_h}\gg 1$, $\gamma_{\varpi_h} \not \ll 1$, $\gamma_{E_h}\gamma_{\varpi_h} \gg 1$), it follows from (\ref{eq:kappap_E0plusEh}) that 
 \begin{align}
\kappa^\prime_E \simeq -5/6+[ 1/6 - (2/3) \gamma^{-2}_{\varpi_h} + 4\gamma^{-4}_{\varpi_h} ] \gamma^{-2}_{E_h} 
\end{align}
reducing to the lower bound $-5/6$ for a purely deterministic $E_s(\tau)$ \cite[sec. III-C]{arnaACFSDF_pt1}.
Dependencies for the cases of quasi-DC ($\gamma_{\varpi_h}\ll 1$) or weak harmonics ($\gamma_{E_h}\ll 1$) follow as in sec. \ref{sec:stirnoisekurt}.

Finally, the combination of noise plus EMI is analyzed in \cite{arnaATRASC2024}, exhibiting additional cross-coupling terms between stir noise and EMI in the expressions for $\cB^{\prime}_E$ and $\kappa^{\prime}_E$. 

\section{Experimental Results}
\subsection{Measurement Configuration}
\begin{figure}[htb] 
\begin{center}
\begin{tabular}{c}
\includegraphics[scale=0.5]{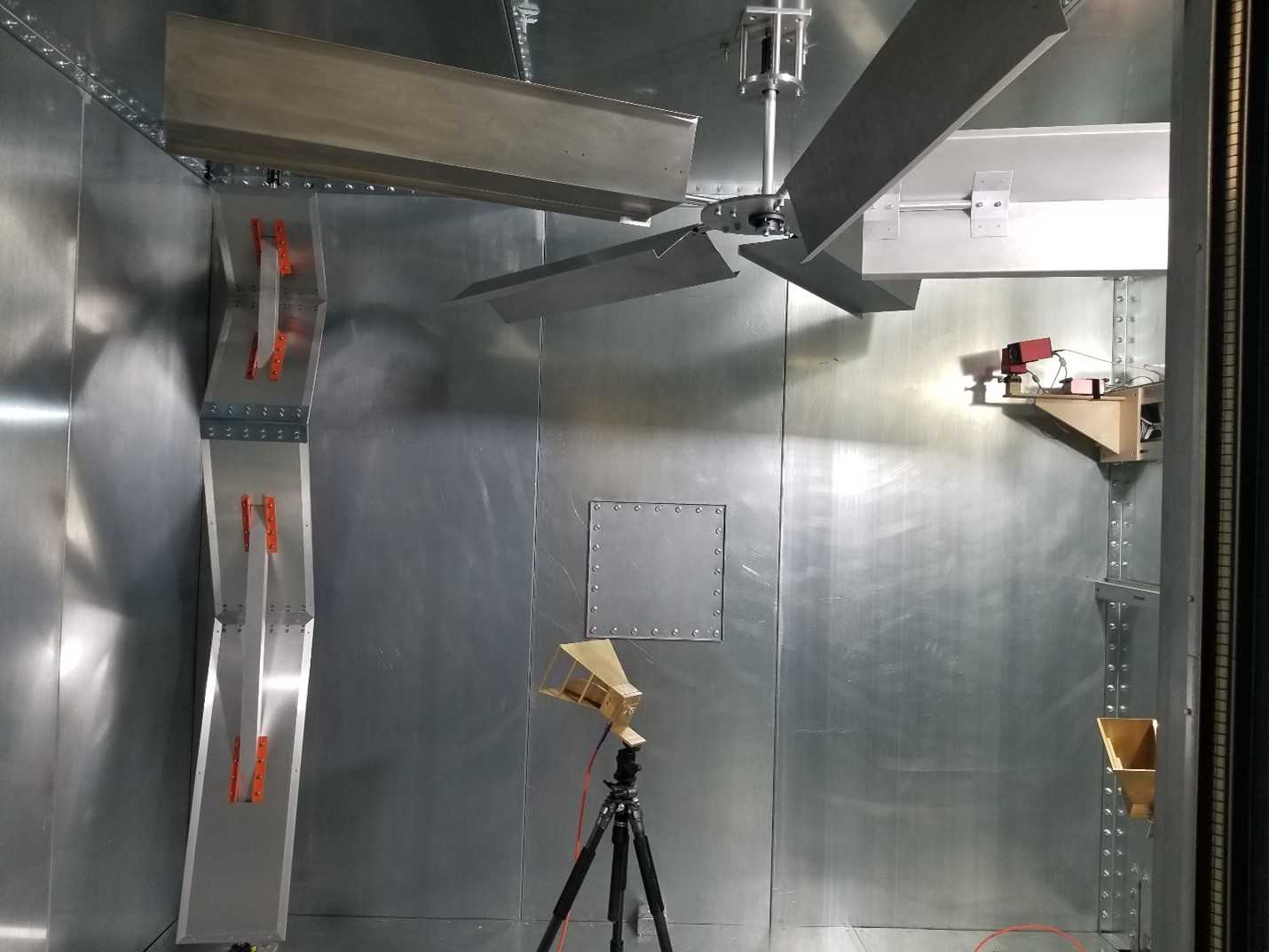}\\ 
\vspace{-1cm}\\
\\
\end{tabular}
\end{center}
{
\caption{\label{fig:NIST_B3}
\small
{
Measurement set-up in the chamber B3 at NIST.
}
}
}
\end{figure}

Measurements were performed inside the  reverberation chamber B3 at NIST, Boulder, CO, with an interior volume of $2.7 \times 3.3\times 2.8$ m$^3$ (Fig. \ref{fig:NIST_B3})
using a $32\,001$-point sampling VNA (model Keysight PNA-L). 
The chamber is furnished with two rotating mode tuners/stirrers: (i) a ``large'' short 
wide five-blade paddle wheel suspended from the ceiling, whose blades are oriented at different randomly chosen angles, generating a swept volume of approximately $1.696$ m$^3$; and (ii) a ``small'' tall slender 
paddle wheel  
stretching from floor to ceiling, with a swept volume of approximately $0.391$ m$^3$.
Both paddles can be operated in tuned (stepped) or stirred (continuous) mode of rotation. For most results reported here, and unless stated otherwise, the small paddle is used in tuned mode, generating $N_t=72$ tune states with equiangular separation of $5$ deg and observing a one-minute settling time after each tuner rotation step, whereas the large paddle is operated in stirred mode, sampled at $N_s=29\,869$ time points across one full stirrer rotation (cf. sec. \ref{sec:period_drift}).  
Each one of two dual ridged horn antennas is directed towards one paddle, without any intentional line-of-sight direction of coupling.
An amplifier (rated 2--18 GHz) was connected at the transmitter side, 
to mitigate secondary noise effects generated by the amplifier affecting the received signal. 
Collected $S_{21}(t)$ data were not normalized by $1-S_{11}(t)$ because $S_{11}(t)$ cannot be measured simultaneously in a MSRC, while synchronization with $S_{11}(t)$ data measured during a subsequent rotation suffers from potential misalignment, jitter and drift. 
Typical VNA-based measurement systems exhibit lower sampling rates but higher SNR levels than vector spectrum analyzer (VSA) systems, so that the effects of level and bandwidth of noise on the bias and variance of the spectral moments can be mitigated.

\subsection{ASDF/ACF Uncertainty Relationship\label{sec:SDFuncert}}
The selected IFBW $B$ and the corresponding minimum time interval $T$ needed to capture $N$ samples 
(dwell time $\Delta \tau = T/N$ per sample) 
are related via \cite{bend1986}
\begin{align}
B \times (T/N) \geq \pi
.
\label{eq:IFBW_uncertainty}
\end{align}
For the $[0/1]$-order 
$g^\prime_E(\varpi)$ \cite[eq. (34)]{arnaACFSDF_pt1}, the lower limit in (\ref{eq:IFBW_uncertainty}) is reached for the bandwidth and correlation length   \cite{bend1986}
\begin{align}
B &= \frac{\int^{\infty}_0 g^\prime_E(\varpi) \rmd \varpi}{{\rm max}_\varpi \{g^\prime_E(\varpi)\} } = \sqrt{\frac{\lambda^\prime_2}{2}}\\
\tau_{c} &= \frac{\int^{+\infty}_{-\infty} |\rho^\prime_E(\tau)| \rmd \tau}{\rho^\prime_E(0)} = \pi \sqrt{\frac{2}{\lambda^\prime_2}}
\end{align}
respectively, such that $\tau_c=\Delta\tau$, $g^\prime_E(B) = \sqrt{2/\lambda^\prime_2} / \rme$ and $\rho^\prime_E(\tau_{c}) = 1/(1+\pi^2)$. 
For $U=|E|^2$, the [0/1] model \cite[eqs. (73) and (76)]{arnaACFSDF_pt1} produces the same lower limit as in (\ref{eq:IFBW_uncertainty}).

\begin{figure}[htb] 
\begin{center}
\begin{tabular}{c}
\vspace{-0.7cm}\\
\hspace{-0.6cm}
\includegraphics[scale=0.65]{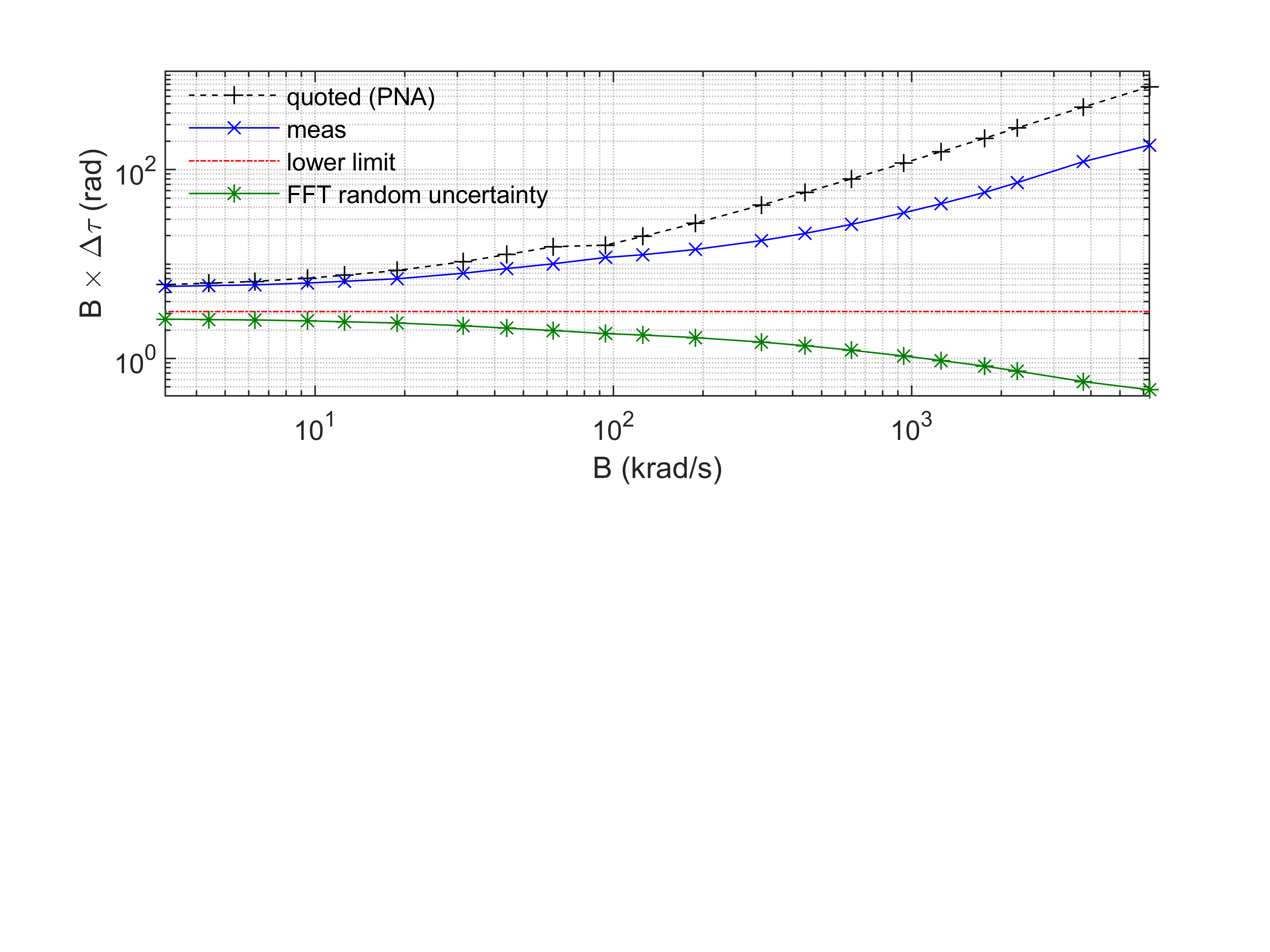}\\
\\
\vspace{-4.5cm}\\
\end{tabular}
\end{center}
{
\caption{\label{fig:SDF_uncertainty}
\small
{
Manufacturer quoted (theoretical) and measured bandwidth--time products (in units rad), its lower limit (\ref{eq:IFBW_uncertainty}) for $[0/1]$-order model, and standardized random uncertainty for FFT-based SDF as a function of IFBW, $B$ (in units krad/s). 
}
}
}
\end{figure}

For an experimental verification of (\ref{eq:IFBW_uncertainty}), a different VNA was employed, capable of collecting $100\,003$ samples across one rotation, thus allowing for slower stirring without increasing $\Delta \tau$. 
Fig. \ref{fig:SDF_uncertainty} compares results quoted by the manufacturer 
against our measured data inside the MSRC.
The estimated uncertainty for the SDF computed via FFT,  from sampled data or ACF, viz., $\sqrt{2\pi N/(B  T)}$ \cite[eq. (8.157)]{bend1986}, is also plotted.
In the remainder, $B = 2\pi \times 10$ krad/s will be selected as the nominal IFBW for all measurements whence $B\times \Delta\tau \simeq 16.7$, as a compromise between measurement duration vs. accuracy.

The lower limit in (\ref{eq:IFBW_uncertainty}) is fundamental to the achievable uncertainty in estimating the SDF or ACF in stir sweeps and, hence, for the spectral moments. 
A narrow IFBW achieves a small bias but yields a large variance of spectral amplitudes, and vice versa. 
Reducing the IFBW necessitates a longer dwell time or larger correlation between samples, requiring a longer time $T$ to complete the stir sweep. 
Thus, for a preset $T$, both temporal and spectral methods of estimating $\lambda_m$ incur a bias that can be decreased by reducing $B$ and increasing $N$.

\subsection{Quasi-Periodicity and Drift \label{sec:period_drift}}
In an ideal MSRC, the start and end points of one mechanical rotation period match perfectly. In practice, electrical and mechanical drift cause a mismatch in the field data after each rotation and raises uncertainty about the true stir period of $E(\tau)$. 
Underestimation of the true period causes a suboptimal exploitation of the stir process, while overestimation results in overlapping data reducing statistical accuracy. 
Hence the stir period should be estimated conservatively. 
Improved control of start and stop times can be achieved using triggering. 

In the presence of long-term drift, the centering of data is an issue because averages dependent on the length $T$, such that averaging over one stir period $T_s$ does not necessarily coincide with averaging over longer or shorter $T$. For lack of a better alternative, the drift is assumed to be constant during one full rotation.

The stir period $T_s$ was estimated by cross-correlating the data with a contiguous subset of duration $T_w\ll T_s$ using a rectangular window. 
The uncertainty on the extracted number $N_s$ was found to be a weak function of the choice of $T_w$ and estimated to be on the order of $0.03\%$.  
The number of nonoverlapping samples is, thus, estimated to be $N_s = 29\,869\pm 9$, constituting one stir period $T_s = 7.9340\pm0.0024$ s, extracted from a stir sweep of $N=32\,001$ samples over $T=8.5$ s that includes an arc of partial overlap spanning $\pi/15$ rad. 
The sampling time resolution is then $\Delta\tau = T_s/N_s \simeq 265.625$ $\mu$s; the corresponding sampling spectral resolution (angular stir speed) is $\Delta \varpi = 2\pi/(N_s\Delta\tau) \simeq 0.79194\pm 0.0003$ rad/s. 

\subsection{Estimation of Spectral Parameters}
\subsubsection{Effect of Field Noncircularity\label{sec:eff_finitediff}}
As noted in sec. \ref{sec:FDdef}, unequal orders of differentiation $n$ and $p$ for $E(\tau)$ and $E^*(\tau)$ may lead to discrepancies in the estimated value of $\lambda_m$ when $E(\tau)$ is noncircular.
Tbl. \ref{tbl:specmom2mom3} illustrates the effect of different choices on the first four 
moments extracted from $S_{21}(\tau)$, denoted here as $E(\tau)$, at $f=18$ GHz. 
In the remainder, the calculated $\lambda_m$ use the complex field instead of individual I- and/or Q-components. 
In other words, $\lambda_0 = \langle | {E} (\tau) |^2\rangle $, $\lambda_1 =  -{\rm Im}[  \langle {E}(\tau) \dot{E}^*(\tau) \rangle ]$, $\lambda_2 = \langle |\dot{E}(\tau)|^2\rangle$, $\lambda_3 = - {\rm Im} [ \langle \dot{E}(\tau) \ddot{E}^* (\tau) \rangle ]$, etc., are selected. 

\def\arraystretch{1.4}
\begin{table}
\begin{center}
\begin{tabular}{||l|l||l|l||}\hline\hline
$\lambda_0 / 2$ & $-$ & $\lambda_1 / 2$ & rad/s\\ \hline\hline
\hspace{-2.5mm} $\langle [ {E}^\prime (\tau) ]^2\rangle$ & $3.91 \times 10^{-5}$ \hspace{-2.4mm} & \hspace{-2.5mm} $\langle {E}^\prime(\tau) \dot{E}^{\prime\prime} (\tau) \rangle$ & $6.03 \times 10^{-4}$ \hspace{-2.4mm} \\ 
\hspace{-2.5mm} $\langle [ {E}^{\prime\prime} (\tau) ]^2\rangle$ & $3.77 \times 10^{-5}$ \hspace{-2.4mm} & \hspace{-2.5mm} $- \langle \dot{E}^\prime(\tau) {E}^{\prime\prime} (\tau) \rangle$ & $6.12 \times 10^{-4}$ \hspace{-2.4mm} \\ 
\hspace{-2.5mm} $\langle | {E} (\tau) |^2\rangle / 2$ \hspace{-1.5mm} & $3.84 \times 10^{-5}$ \hspace{-2.4mm} & \hspace{-2.5mm} $ -{\rm Im}[  \langle {E}(\tau) \dot{E}^*(\tau) \rangle ] / 2$ \hspace{-1.5mm} & $6.08 \times 10^{-4}$ \hspace{-2.4mm} \\ \hline\hline
\end{tabular}
~\vspace{0.2cm}\\
\begin{tabular}{||l|l||l|l||}\hline\hline
$\lambda_2 / 2$ & (rad/s)$^2$ \hspace{-2mm} & $\lambda_3 / 2$ & (rad/s)$^3$ \hspace{-3mm}\\ \hline\hline
\hspace{-2.5mm} $\langle [ \dot{E}^\prime (\tau) ]^2\rangle$ & $13.75$ & \hspace{-2.5mm} $\langle \dot{E}^\prime(\tau) \ddot{E}^{\prime\prime} (\tau) \rangle$ & $615.5$ \\ 
\hspace{-2.5mm} $\langle [ \dot{E}^{\prime\prime} (\tau) ]^2\rangle$ & $13.48$ & \hspace{-2.5mm} $- \langle \ddot{E}^\prime(\tau) \dot{E}^{\prime\prime} (\tau) \rangle$ & $626.9$\\ 
\hspace{-2.5mm} $\langle | \dot{E} (\tau) |^2\rangle / 2$ & $13.61$ & \hspace{-2.5mm} $- {\rm Im} [ \langle \dot{E}(\tau) \ddot{E}^* (\tau) \rangle ] / 2$ \hspace{-2mm} & $621.2$ \\ \hline
\hspace{-2.5mm} $- \langle E^\prime(\tau) \ddot{E}^{\prime} (\tau) \rangle$ & $12.73$ & \hspace{-2.5mm} $-\langle {E}^\prime(\tau) \dddot{E}^{\prime\prime} (\tau) \rangle$ & $468.1$ \\ 
\hspace{-2.5mm} $- \langle {E}^{\prime\prime}(\tau) \ddot{E}^{\prime\prime} (\tau) \rangle$ & $12.45$  & \hspace{-2.5mm} $\langle \dddot{E}^\prime(\tau) {E}^{\prime\prime} (\tau) \rangle$ & $476.5$\\
\hspace{-2.5mm} $- {\rm Re} [ \langle {E}(\tau) \ddot{E}^* (\tau) \rangle ] / 2$ \hspace{-2mm} & $12.59$ & \hspace{-2.5mm} $ {\rm Im} [ \langle {E}(\tau) \dddot{E}^* (\tau)\rangle ] / 2$ \hspace{-2mm} & $472.3$\\
\hline\hline
\end{tabular}
\caption{\label{tbl:specmom2mom3} \small 
Alternative discrete-time estimates of non-normalized $\lambda_{0\ldots 3} / 2$ from sampled stir sweeps for large stirrer at $f=18$ GHz. Overdots denote FD; $\tau$ denotes $\tau_n=n\Delta\tau$ for $n=0,1,\ldots,N_s - 1$.}
\end{center}
\end{table}

\subsubsection{Effect of Input Power and Electrical  Noise\label{sec:eff_inputpower}}
In \cite{mill1970b}, \cite{mill1972}, effects of additive white noise on the first two spectral moments of a Gaussian ASDF  were analyzed;
for the effect on a complete exponential ASDF, cf. \cite[sec. VI]{arnaACFSDF_pt1}. 
The effect of noise becomes progressively more significant when $E(\tau)$ fades at one or more stir states, on approaching the noise floor of the VNA or when introducing an external noise source. 

\begin{figure}[htb] 
\begin{center}
\begin{tabular}{cc}
\hspace{-0.3cm}
\includegraphics[scale=0.3]{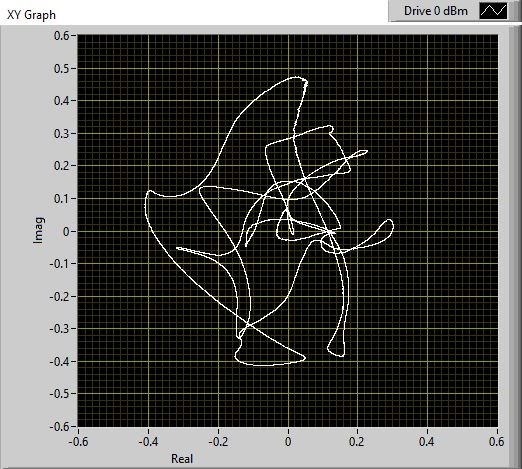} & \includegraphics[scale=0.3]{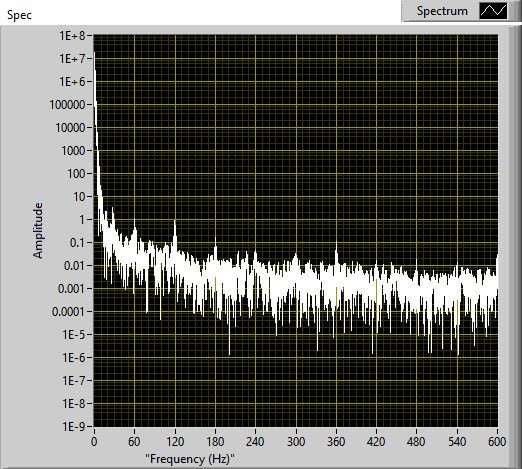}\\
(a) & (b)\\
\hspace{-0.3cm}
\includegraphics[scale=0.3]{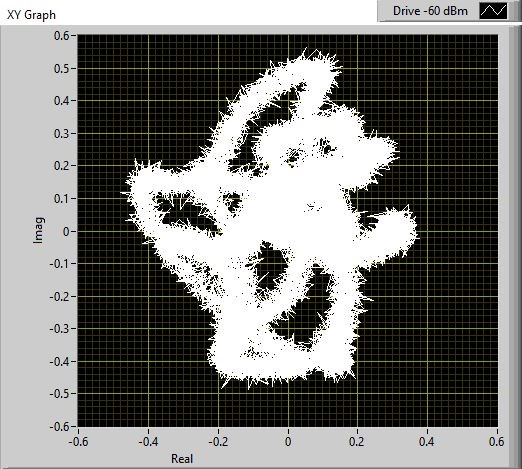} & \includegraphics[scale=0.3]{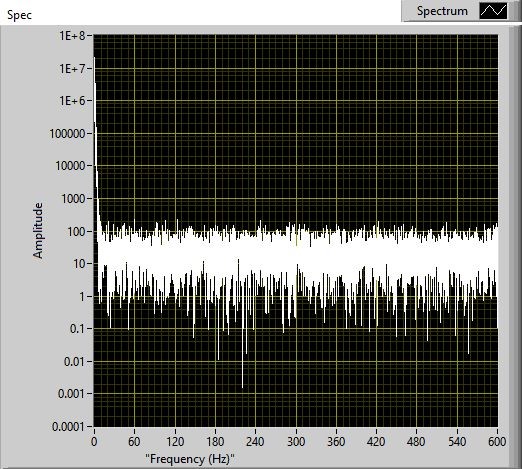}\\
(c) & (d)\\
\end{tabular}
\end{center}
{
\caption{\label{fig:John_scatterplottrace_ifv_Pin}
\small
{
(a)(c) Scatter plots of stir traces $({\rm Re}[S_{21}(\tau)],{\rm Im}[S_{21}(\tau)])$; (b)(d) associated stir ASDFs $g^\prime_{S_{21}}(\varpi)$, for (a)(b) $P_{\rm in}=0$ dBm and for (c)(d) $P_{\rm in}=-60$ dBm, all measured at $f=1$ GHz, without amplifier, using small paddle as stirrer. 
}
}
}
\end{figure}

Figs. \ref{fig:John_scatterplottrace_ifv_Pin}(a) and (c) show the effect of lowering $P_{\rm in}$ on $S_{21}(\tau)$ at $f=1$ GHz without amplifier. 
This reduction of the input power also manifests itself in the corresponding SDFs, shown in Figs. \ref{fig:John_scatterplottrace_ifv_Pin}(b) and (d). These plots confirm that lowering the SNR in a stir trace ($P_{\rm in} = -60$ dBm) results in a whitening of the original SDF at $P_{\rm in} = 0$ dBm.  

\begin{figure}[htb] 
\begin{center}
\begin{tabular}{c}
\hspace{-0.5cm}
\includegraphics[scale=0.65]{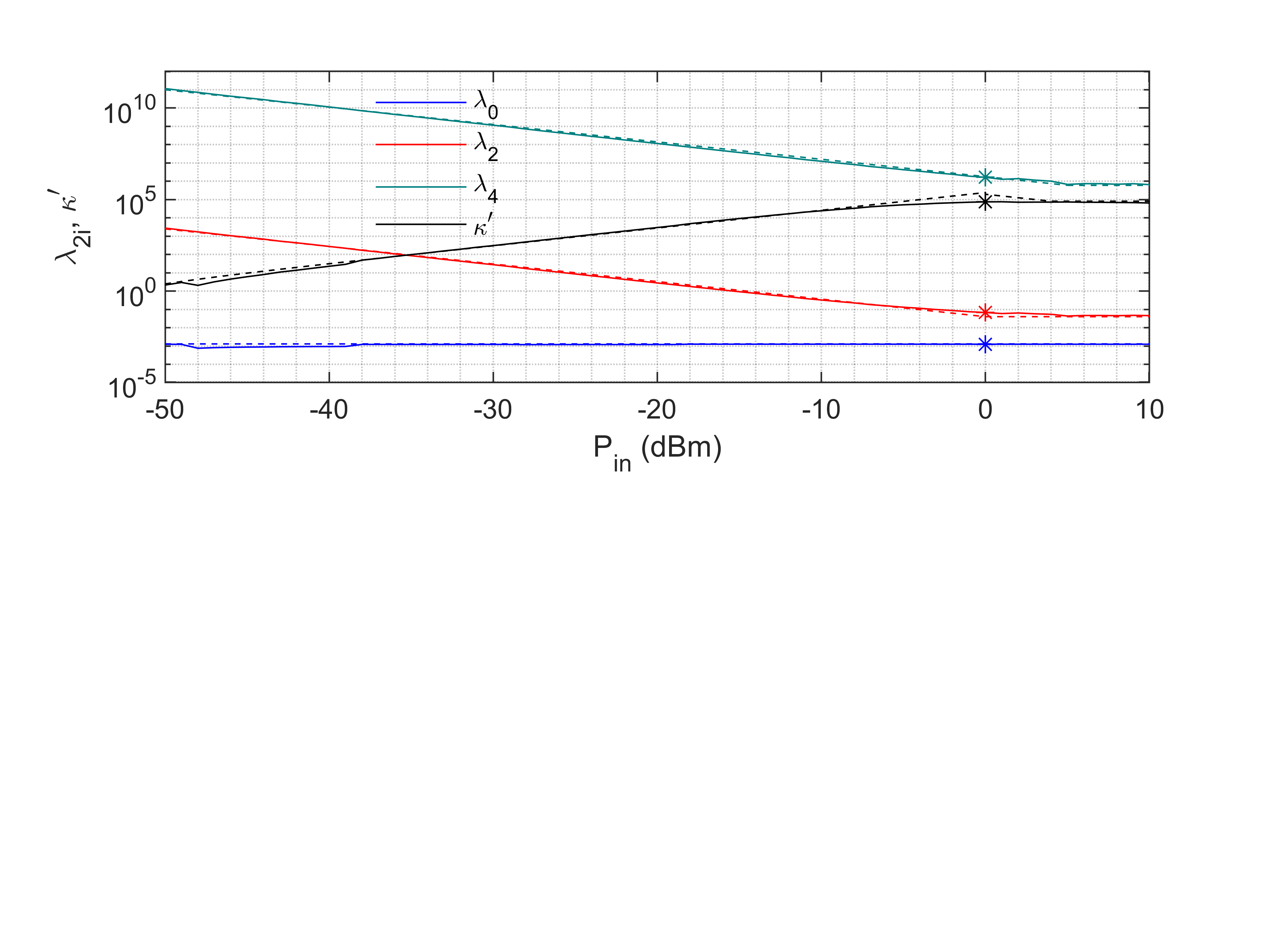}\\
\vspace{-4.7cm}\\
\\
\end{tabular}
\end{center}
{
\caption{\label{fig:John_specmom_ifv_Pin}
\small
{
Experimental values (solid) and first-order logarithmic fits (\ref{eq:model_lambda0_lin})--(\ref{eq:model_kappap_lin}) (dashed) for $\lambda_{2i}$ (in units (rad/s)$^{2i}$) and $\kappa^\prime$ as a function of drive input power $P_{\rm in}$ (in units dBm) at $f=1$ GHz, without amplifier, using small paddle as stirrer}.
Asterisks indicate nominal $P_{\rm in}=0$ dBm.  
}
}
\end{figure}

Fig. \ref{fig:John_specmom_ifv_Pin} shows the experimental non-normalized $\lambda_{2i}(P_{\rm in})$ and $\kappa^\prime(P_{\rm in})$ at $f=1$ GHz when reducing $P_{\rm in}$ in steps of $1$ dBm,
together with first-order logarithmic fits, obtained as
\begin{align}
\lambda_0 (P_{\rm in}) &\simeq 1.288 \times 10^{-3} 
\label{eq:model_lambda0_lin}
\\
\lambda_2 (P_{\rm in}) &\simeq \left \{
\begin{array}{ll}
(5.37 \times 10^{-5}) P^{-0.96}_{\rm in}, &P_{\rm in} \ll 1\,{\rm mW}\\
4.074 \times 10^{-2}, &P_{\rm in} \gg 1\,{\rm mW}
\end{array} \right .
\label{eq:model_lambda2_lin}
\\
\lambda_4 (P_{\rm in}) &\simeq \left \{
\begin{array}{ll}
(2.60 \times 10^{3}) P^{-0.95}_{\rm in}, &P_{\rm in} \ll 3\,{\rm mW}\\
6.310 \times 10^{5}, &P_{\rm in} \gg 3\,{\rm mW}
\end{array} \right .
\label{eq:model_lambda4_lin}
\\
\kappa^\prime (P_{\rm in}) &\simeq \left \{
\begin{array}{ll}
(1.94 \times 10^8) P^{0.97}_{\rm in}-1, &P_{\rm in} \ll 1\,{\rm mW}\\
8.161 \times 10^4,                  &P_{\rm in} \gg 3\,{\rm mW}
\end{array} \right .
\label{eq:model_kappap_lin}
\end{align}
suggesting that, at least to good approximation, 
\begin{align}
\lambda^\prime_2, \, \lambda^\prime_4 \propto ( P_{\rm in})^{-1},
\hspace{0.5cm} 
\kappa^\prime \propto P_{\rm in}
\end{align} 
when $P_{\rm in} \ll 0$ dBm.
For $\kappa^\prime$, this result agrees with the theoretical relationship $|\kappa^\prime_E(\gamma_N)| \propto \gamma^{-2}_N \propto P_{\rm in}/P_N$ for $\gamma_N \gg 1$, constant $P_N$ and $\cB^\prime_N=B$; cf. (\ref{kappap_E0plusN_smallgammaN}) and Fig. \ref{fig:kappap_ifv_SNR}(b).

The findings for $\lambda_0(P_{\rm in})$ and $\lambda_2(P_{\rm in})$ agree\footnote{In \cite{mill1970b}, explicit numerical results apply to the case of Gaussian SDFs.} with the theoretical result that the bias of $\lambda^\prime_2(P_{\rm in})$ increases rapidly with decreasing SNR \cite[sec. V.A.]{mill1970b}. When the spectral width of the stir process is much larger than the stir spectral resolution, this increase happens at a rate of approximately $1/P_{\rm in}$.  
Thus, noise induces a bias for estimated $\lambda^\prime_{2i}$; for $\gamma_N \ll 1$ (cf. (\ref{eq:lambdan_E0plusN}))
\begin{align}
\lambda^{\prime}_{m} \simeq 
{m!}{(\cB^\prime_s)^{m}} \left [ 1 + \left ( \frac{\gamma^{m}_B}{(m+1)!} - 1 \right ) \gamma^{2}
_N\right ]
.
\end{align}

\subsection{Comparison of Stir-Domain Estimation Methods\label{sec:sweep_vs_corr}}
Tbl. \ref{tbl:sweep_vs_corr} compares estimates of $(\lambda^{\prime(\prime)}_m)^{1/m}$ obtained from the sweep vs. ACV FD methods, indicating close numerical agreement. 
Associated centered moment-based quantities are similarly close; e.g., the relative difference of $[\lambda^\prime_2-(\lambda^{\prime\prime}_1)^2]^{1/2}$ between both methods is less than $1.4 \times 10^{-4}$. As functions of ratios of raw moments and $\lambda_0$, these estimates are still subject to sample statistical uncertainty \cite[sec. 27.7]{cram1946}.
Compensation for FD bias yields a positive and smaller (in magnitude) $\kappa^\prime$ for both methods, whereas $\kappa^{\prime\prime}$ becomes closer to zero but remains negative for the stir sweep method, unlike for ACV.
FD debiasing does not remove the overall bias, as other contributions including aliasing, finite sample sweep length, etc., remain 
\cite{zrni1977},
but are relatively small ($\sim 1/N_s$). 

From $(\lambda^{\prime(\prime)}_m)^{1/m}$, several spectral statistics follow, including the mean $\mu_{g^{\prime(\prime)}_{E}} = \lambda^{\prime(\prime)}_1$, variance $\sigma^2_{g^{\prime(\prime)}_{E}} = \lambda^{\prime(\prime)}_2-(\lambda^{\prime(\prime)}_1)^2$, 
and normalized bandwidths \cite{long1956}, \cite{vanm1972}
\begin{align}
\delta^{\prime(\prime)}_E 
&\equiv \sigma_{g^{\prime(\prime)}_{E}} / \sqrt{\lambda^{\prime(\prime)}_2} = \sqrt{1-\left ( \lambda^{\prime(\prime)}_1 \right )^2/\lambda^{\prime(\prime)}_2}\\
\delta^{\prime}_E &\simeq \sqrt{ \frac{1}{2}} \left [ 1 - \frac{1}{4} \left ( \frac{\Delta\tau}{\beta^{\prime}} \right )^2 \right ]
\label{eq:specBW_delta_approx}\\
\epsilon^{\prime(\prime)}_E &\equiv \sqrt{1-1/k_{g^{\prime(\prime)}_{E}} } = \sqrt{1-\left  ( \lambda^{\prime(\prime)}_2 \right )^2/\lambda^{\prime(\prime)}_4}\\
\epsilon^{\prime}_E &\simeq \sqrt{ \frac{5}{6} } \left [ 1 - \frac{3}{10} \left ( \frac{\Delta\tau}{\beta^{\prime}} \right )^2 \right ]
.
\label{eq:specBW_eps_approx}
\end{align}
The approximations (\ref{eq:specBW_delta_approx}) and (\ref{eq:specBW_eps_approx}) account for FD bias, which is again negative and quadratic to leading order. 
In Tbl. \ref{tbl:sweep_vs_corr}, the FD compensated $\delta^{\prime\prime}_E$ at $f=18$ GHz is within $0.2\%$ of
its maximum value $1$, while $\delta^\prime_E$ and $\epsilon^\prime_E$ also remain marginally below their respective theoretical minimum values, $\sqrt{1/2}$ and $\sqrt{5/6}$, for ideal I/Q isotropy in the complex $E$-plane, i.e., for circular local fields \cite{arnalocavg}. This indicates a wide stir (dynamic) bandwidth despite the narrow excitation (static) bandwidth.
\def\arraystretch{1.4}
\begin{table}
\begin{center}
\begin{tabular}{||l|r|r||}\hline\hline
Parameter & Sweep FD & ACV FD \\ \hline\hline
$\lambda^{\prime\prime}_1$ & $15.8299$  & $16.0401$\\ 
$(\lambda^{\prime}_2)^{1/2}$ & $595.542$  & $595.630$  \\ 
$(\lambda^{\prime\prime}_3)^{1/3}$ & $252.940$  & $257.284$  \\ 
$(\lambda^{\prime}_4)^{1/4}$ & $926.161$   & $926.478$ \\ 
$(\lambda^{\prime\prime}_5)^{1/5}$ & $550.232$  & $572.812$ \\ \hline\hline
$\kappa^\prime$ (biased)& $-0.02514$  & $-0.02437$ \\ 
$\kappa^\prime$ (FD compensated)& $+0.01118$  & $+0.01195$ \\ \hline
$\kappa^{\prime\prime}$ (biased) & $-0.08541$  & $+0.02307$ \\
$\kappa^{\prime\prime}$ (FD compensated) & $-0.04193$  & $+0.06804$ \\ \hline
$\delta^\prime_E$ (biased) & $0.69542$ & $0.69533$ \\ 
$\delta^\prime_E$ (FD compensated)& $0.70494$ & $0.70485$ \\ \hline
$\delta^{\prime\prime}_E$ (biased) & $0.99965$ & $0.99964$ \\ 
$\delta^{\prime\prime}_E$ (FD compensated)& $0.99805$ & $0.99811$ \\ \hline
$\epsilon^\prime_E$ (biased) & $0.90704$ & $0.90711$ \\ 
$\epsilon^\prime_E$ (FD compensated)& $0.91051$ & $0.91059$ \\ 
\hline\hline
\end{tabular}
\caption{\label{tbl:sweep_vs_corr} \small $(\lambda^{\prime(\prime)}_m)^{1/m}$ for $1\leq m \leq 5$, in units rad/s; $\kappa^{\prime(\prime)}$; $\delta^{\prime(\prime)}_E$; $\epsilon^{\prime}_E$, extracted using stir sweep vs. ACV FD methods, at $f=18$ GHz, for an arbitrary secondary tune state, with $\lambda_0 = 7.6774 \times 10^{-5}$.}
\end{center}
\end{table}

\subsection{SDF Symmetry, Phase Linearity and Doppler Shift}
WSS $E^\prime(\tau)$ and $E^{\prime\prime}(\tau)$ have Hermitean ASDFs $g_{E^\prime}(\varpi)$ and $g_{E^{\prime\prime}}(\varpi)$.
This complex conjugate symmetry does not necessarily extend to $g_E(\varpi)$ for sampled fields \cite{mill1970b}, \cite{pass1983}. An asymmetric ASDF corresponds to a nonlinearity of $\phi_{\rho_E}(\tau)$ that contributes to 
 (\ref{eq:lambda3_complexACFmethod})--(\ref{eq:lambda5_complexACFmethod}) and, hence, to spectral skewness. Such phase discrimination can be used, e.g., for characterizing narrowband interference \cite{ward2014}. 

The continuous motion of a mechanical stirrer carries an associated Doppler shift (mean velocity and $\langle \Lambda_1 \rangle \not = 0$), raising the question of significance of its effect on $g_{E}(\varpi)$, $\delta^{\prime(\prime)}_E$ and $\kappa$. To this end, a perturbation expansion\footnote{Applicability of the perturbation expansion method requires $\rho^2_E(\Delta\tau) \gg 1/N_s$ and $\Delta\tau/\beta \gg 1/N_s$ \cite{zrni1979}. 
} 
of the complex ACF of $E(\tau)$ can be applied \cite{zrni1979}.
If $N_e(\tau)$ is negligible, then the variance of $\Lambda_1 / \Lambda_0$ for $\rho_E(\Delta\tau) = [1 + (\Delta\tau/\beta)^2]^{-1}$ is estimated as
\begin{align}
\sigma^2_{\Lambda_1/\Lambda_0} \simeq \frac{\pi}{2(\Delta\tau)\beta N_s}
\left [ 1 + \frac{1}{2} \left ( \frac{\Delta\tau}{\beta} \right )^2 \right ]
.
\label{eq:asymp_sigmasq_Lambda1}
\end{align}
At $f=18$ GHz, this yields $\sigma_{\Lambda_1/\Lambda_0} \simeq 9.2$ rad/s, i.e., comparable in magnitude to 
$\langle \Lambda_1 / \Lambda_0 \rangle_{N_t} \simeq 21.22$ rad/s, but about two orders of magnitude smaller than $\sigma_{g_E}$. 
On this basis, the assumption of a quasi-stationary symmetric ASDF with negligible stir frequency shift is justifiable.

Physically, a significantly nonzero value of $\langle \Lambda_1  / \Lambda_0 \rangle_{N_t}$ can be associated with a net mean radial flux of directed energy produced by reflections from the stirrer's surface elements across the stir process. 
This flux arises most prominently when the antenna's boresight direction is pointing towards an approaching or receding paddle blade, i.e., directed off-axis with reference to the stirrer's shaft. The spectral width $\langle(\Lambda_2-{\Lambda^2_1})/\Lambda_0\rangle^{1/2}_{N_t}$ then measures the RMS spread of the instantaneous radial flux during the stir process, serving as a measure of stir efficiency.
Higher-order spectral moments can be used to further refine the probability distribution of this flux.

\subsection{Effect of Sampling Rate and Decimation\label{sec:eff_decim}}
\begin{figure}[htb] 
\begin{center}
\begin{tabular}{c}
\hspace{-0.6cm}
\includegraphics[scale=0.66]{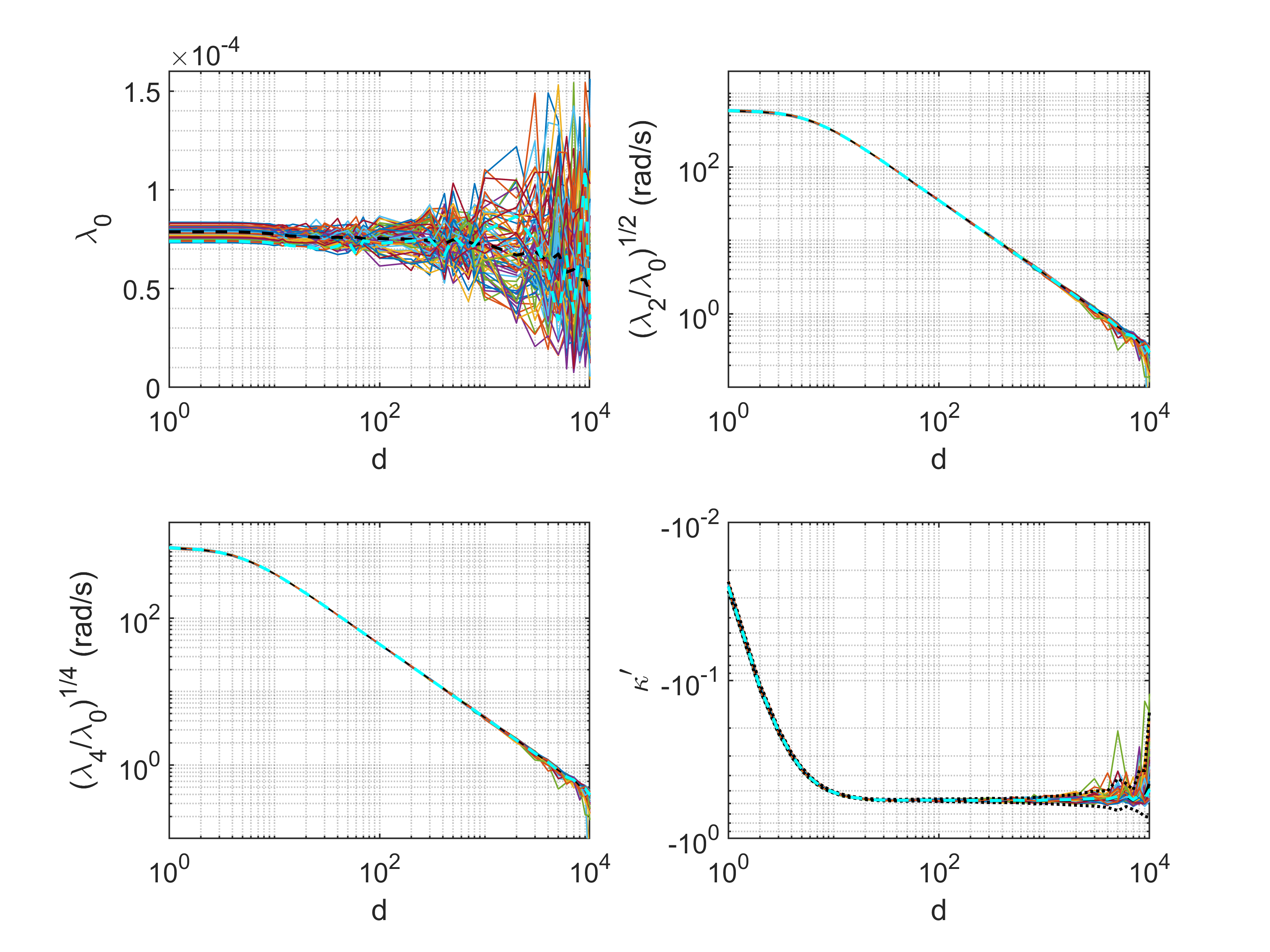}\\
(a)\\
\hspace{-0.6cm}
\includegraphics[scale=0.66]{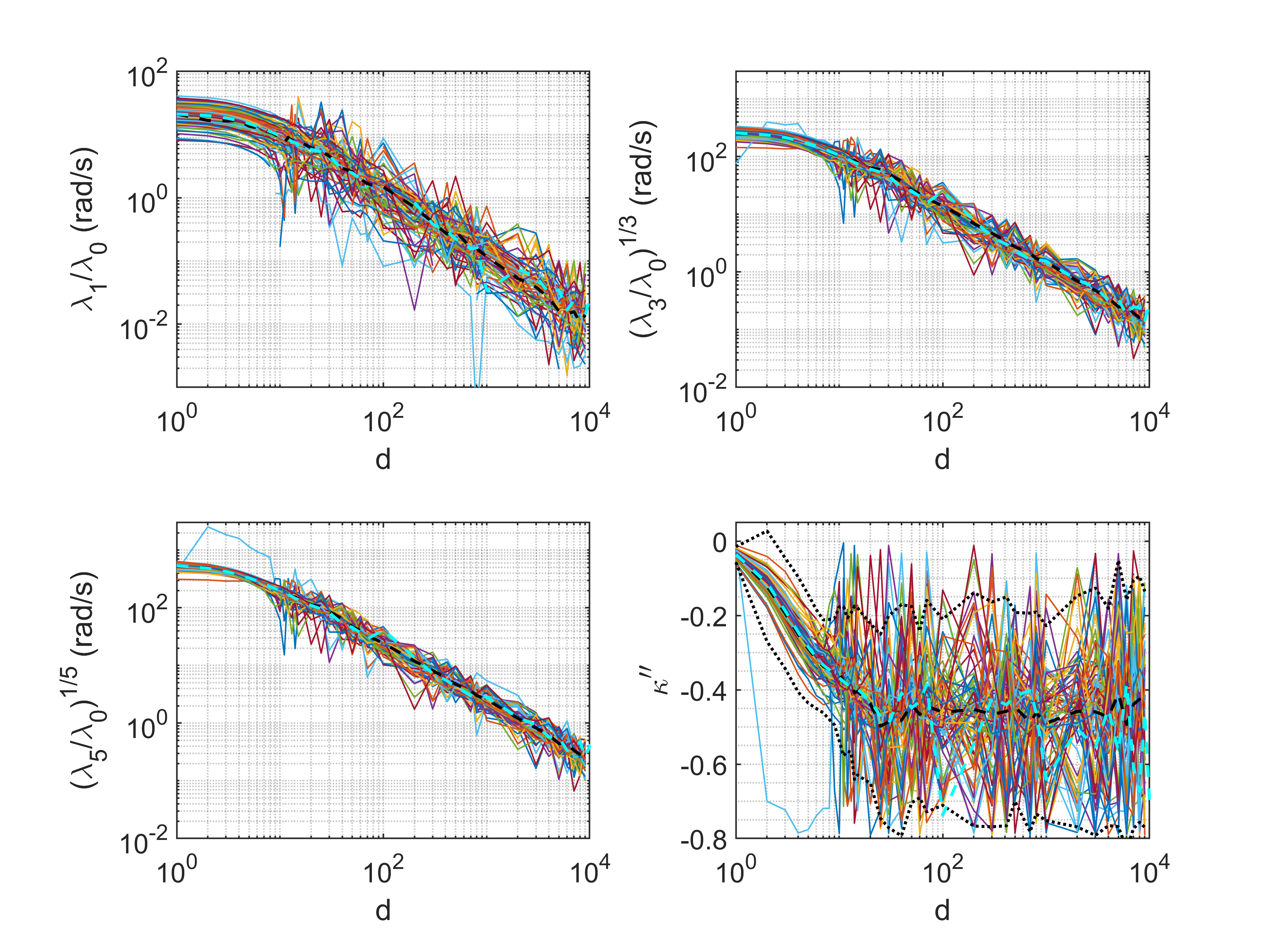}\\
(b)
\end{tabular}
\end{center}
{
\caption{\label{fig:decimation_withamp_f18GHz_traces}
\small
{(a) Even- and (b) odd-order normalized spectral moments $(\lambda_m/\lambda_0)^{1/m}$ and $\kappa^{\prime(\prime)}$ as a function of data decimation factor $d$, based on: 
(i) individual stir sweeps for 72 tune states (solid), 
(ii) averaging across 72 tune states (black dashed), 
(iii) concatenation of all $N_t N_s = 2\,150\,568$ samples 
(cyan dashed),
including confidence intervals $[\langle \kappa^{\prime(\prime)} \rangle - 2 \sigma_{\kappa^{\prime(\prime)}}, \langle \kappa^{\prime(\prime)} \rangle + 2 \sigma_{\kappa^{\prime(\prime)}}]$ (black dotted) at $f=18$ GHz,
for stirring by large paddle, secondary tuning by small paddle, including amplifier.}
}
}
\end{figure}

The effect of sampling on the spectral moments can also be investigated {\it a posteriori}, by subsampling the original data with a decimation factor $d=\tau / \Delta \tau \geq 1$
\cite{arnathresh}.
A continuous stir trace corresponds to the limit $\Delta \tau \rightarrow 0$ and $d=1$.
Fig. \ref{fig:decimation_withamp_f18GHz_traces} shows experimental values of $\lambda_0$, $(\lambda^{\prime(\prime)}_{m})^{1/m}$ and $\kappa^{\prime(\prime)}$ as functions of $d$ at $f=18$ GHz for the large stirrer, obtained using the stir sweep method. 
Expanded confidence intervals $[\langle \kappa^{\prime(\prime)} (d) \rangle - 2 \sigma_{\kappa^{\prime(\prime)}}, \langle \kappa^{\prime(\prime)} (d) \rangle + 2 \sigma_{\kappa^{\prime(\prime)}}]$ 
are shown with black dots.

For even-order moments and coarse subsampling ($30 < d <1000$), the slope of the mean $\langle(\lambda^\prime_m(d))^{1/m}\rangle_{N_t}$ for $m=2$ and $4$ in Fig. \ref{fig:decimation_withamp_f18GHz_traces}(a) is $-0.9991$ and $-0.9995$, respectively, in good agreement with (\ref{eq:estlambdapeven_cont_fd}) for $\Delta\tau/\beta^\prime \gg 1$, to leading order.
For $\Delta\tau/\beta^\prime \ll 1$, the quadratic dependence of $\langle(\lambda^\prime_m(d))^{1/m}\rangle_{N_t}$ in (\ref{eq:estlambdapeven_cont_fd_approx}) is also apparent.
The asymptotic value of $\kappa^\prime(d\rightarrow N_s/2) \simeq -0.584$ is slightly above the theoretical value $-3/4$ in (\ref{eq:bias_kappap_fd_cont}), valid for ideal stirring in the absence of noise, thus indicating good stir performance. 
For $d\rightarrow 1$, the slope of $\log(|\kappa^\prime|)$ vs. $\log(d)$ is sensitive to the specific small value of $|\kappa^\prime|$ at $d=1$, as expected, ranging between $-1.76$ and $-2.49$ across the 72 secondary tune states $\tau_{2,\ell}$. This range incorporates the theoretical value $-2$ for $\Delta\tau/\beta^\prime \ll 1$ from (\ref{eq:bias_kappap_fd_cont}), applicable to ideal $\kappa^\prime(\Delta\tau/\beta^\prime\rightarrow 0)=0$.

For odd-order moments and $\kappa^{\prime\prime}$ shown in Fig. \ref{fig:decimation_withamp_f18GHz_traces}(b), their dependence on $d$ exhibits considerably larger fluctuations across tune states than for even-order moments. 
For $100 < d < 1000$, the slope of $\langle(\lambda^{\prime\prime}_m(d))^{1/m}\rangle_{N_t}$ for $m=1$, $3$ and $5$ is $-1.0372$, $-0.9978$ and $-0.9981$, respectively.
The asymptotic value $\kappa^{\prime\prime}(d \rightarrow N_s/2)\simeq -0.49$ is again slightly above the theoretical value, now $-16/25$ in (\ref{eq:bias_kappapp_fd_cont}), for ideal noise-free stirring.

The differences in concentration of individual $d$-characteristics between even- vs. odd-order moments in  Figs. \ref{fig:decimation_withamp_f18GHz_traces}(a) and (b) can be traced to the unequal orders of differentiation $n \not = p$ when $m$ is odd. This leads to mismatched FD grids of sampled data points for extracting odd-order moments, resulting in different values of $\lambda_m$ depending on whether $n=p+1$ or $n=p-1$; cf. Tbl. \ref{tbl:specmom2mom3}. 

\subsection{CW Frequency Characteristics}
\subsubsection{Even-Order Moments}
\begin{figure}[htb] 
\begin{center}
\begin{tabular}{c}
\hspace{-0.6cm} 
\includegraphics[scale=0.65]{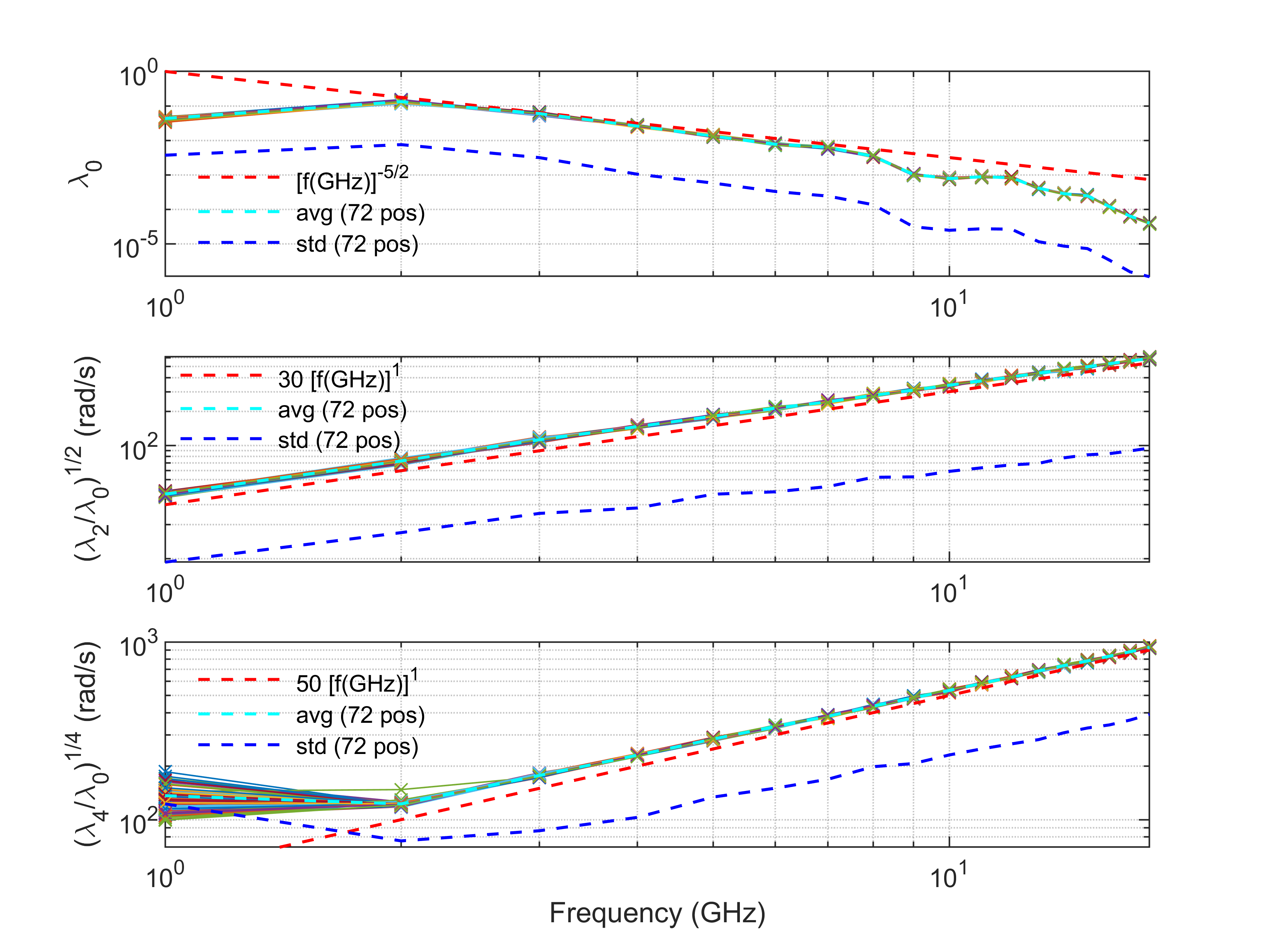}\\ 
\\
\vspace{-2cm}\\
\\
(a)\\
\hspace{-0.6cm}
\includegraphics[scale=0.63]{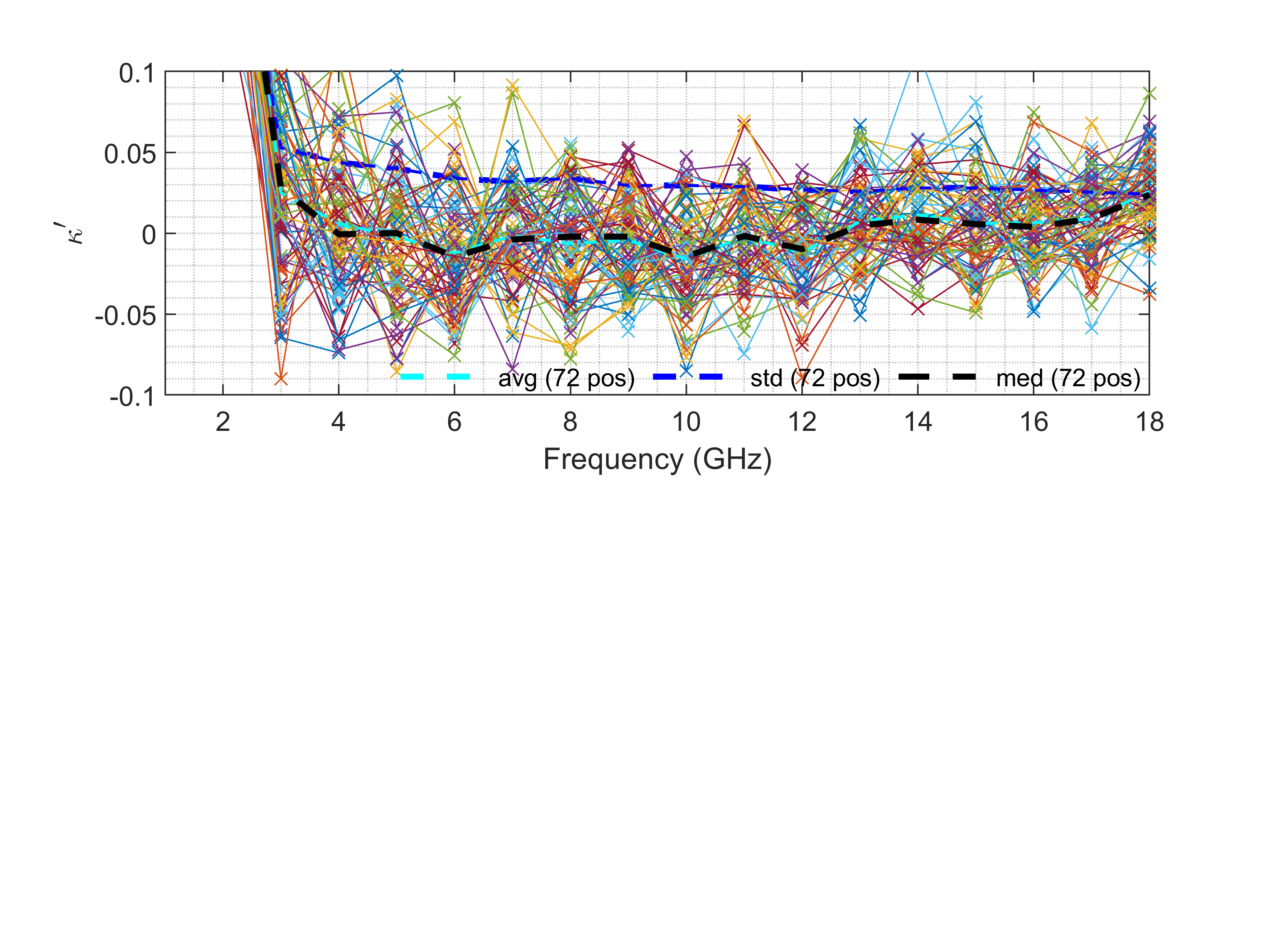}\\ 
\\
\vspace{-5.2cm}\\
\\
(b)\\
\hspace{-0.6cm}
\includegraphics[scale=0.63]{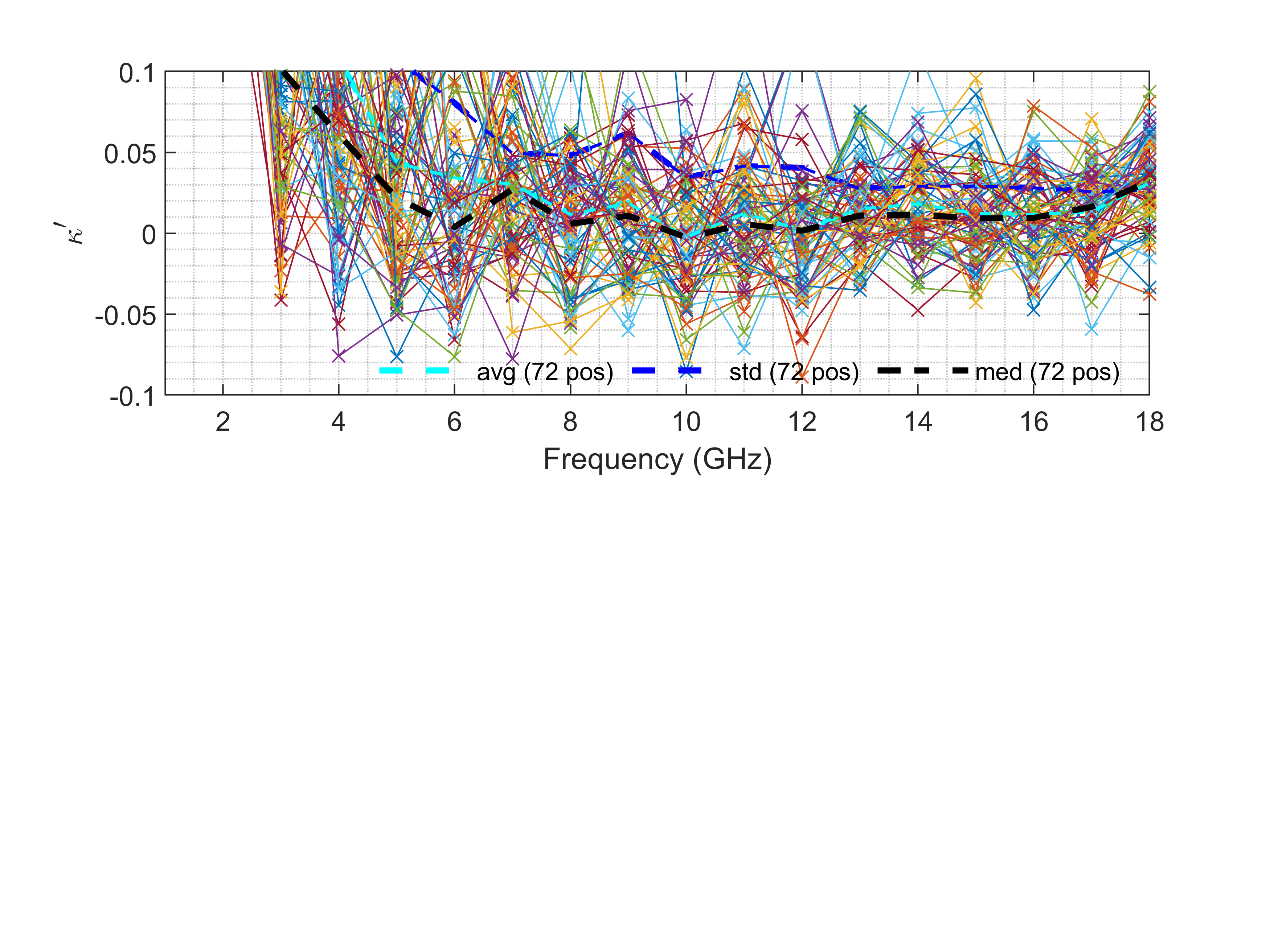}\\ 
\\
\vspace{-5.2cm}\\
\\
(c)
\end{tabular}
\end{center}
{
\caption{\label{fig:evenmoments_f1to18GHz}
\small
{
(a) Normalized even-order spectral moments as a function of CW frequency $f$: 72 individual stir traces (colored, solid), 
tune average (cyan, dashed) and tune standard deviation (blue, dashed), for large stirrer, including amplifier;
(b) FD debiased ${\kappa^\prime}(f)$ based on stir sweep data;
(c) FD debiased ${\kappa^\prime}(f)$ based on complex ACV.
}
}
}
\end{figure}

Fig. \ref{fig:evenmoments_f1to18GHz}(a) shows $\lambda_0(f)$ and $(\lambda_{m}(f)/\lambda_0(f))^{1/m}$ for $m=0,2,4$.
The latter characteristics increase proportionally with $f$ when $d \rightarrow 0$. 
At higher $f$, the curves increasingly sag with increasing $\Delta\tau$; cf. \cite[Fig. 6]{arnathresh}. 

In Fig. \ref{fig:evenmoments_f1to18GHz}(b), ${\kappa^\prime}(f)$ rapidly converges towards ideal zero with increasing $f$, on average, as predicted theoretically. Fig. \ref{fig:evenmoments_f1to18GHz}(b) and (c) show that residual fluctuations  
remain, caused by FD numerical underflow. These can be reduced by taking the RMS value of ${\kappa^\prime}^2$ across the $N_t$ tune states. 
At $f=18$ GHz, this results in $\langle {\kappa^\prime}^2 \rangle^{1/2}_{N_t} \simeq 0.0260$ whereas $\langle {|\kappa^\prime|} \rangle^{1/2}_{N_t} \simeq 0.1448$.

\subsubsection{Odd-Order Moments}
\begin{figure}[htb] 
\begin{center}
\begin{tabular}{c}
\hspace{-0.6cm}
\includegraphics[scale=0.65]{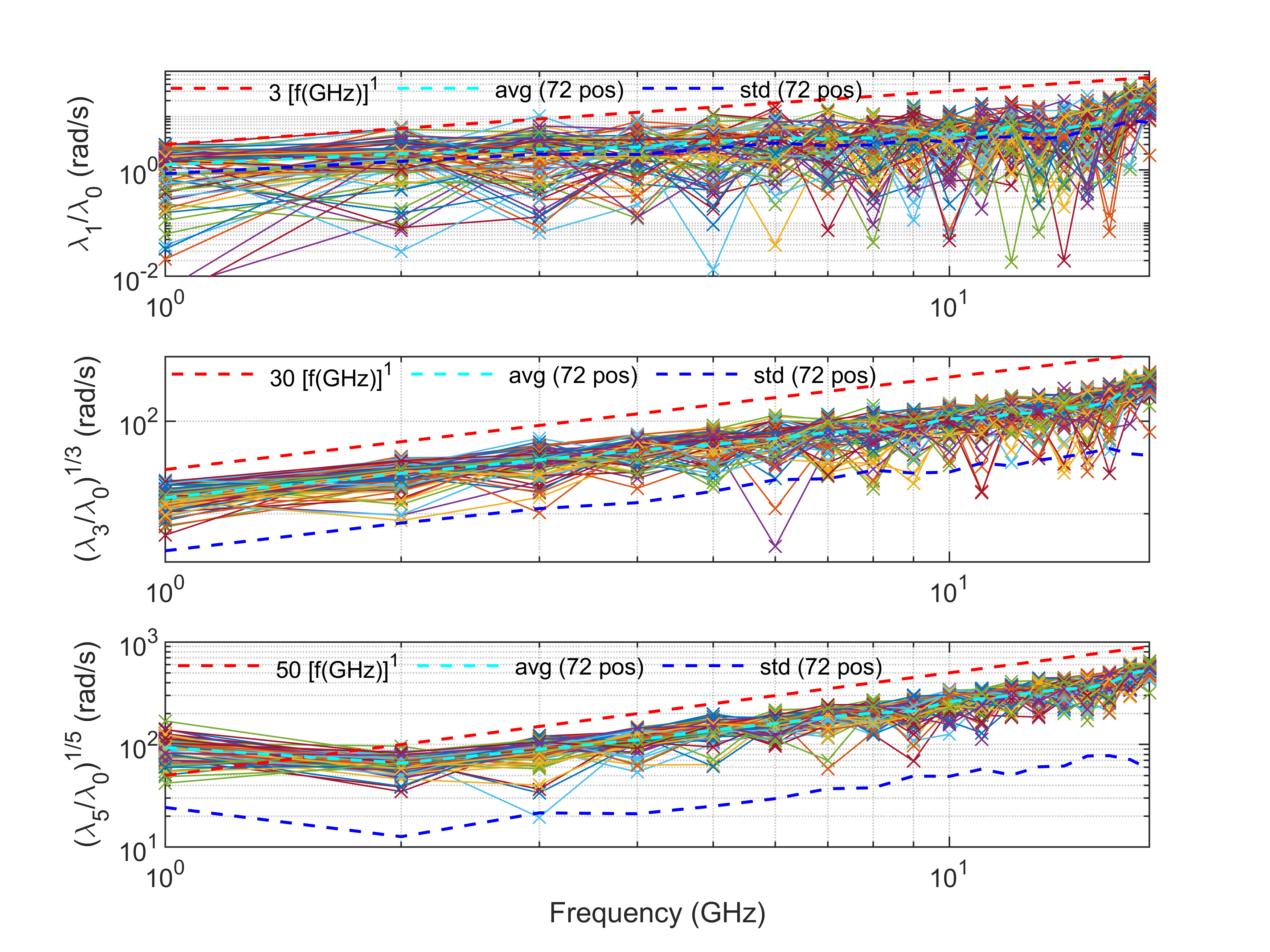}\\ 
\\
\vspace{-2cm}\\
\\
(a)\\
\hspace{-0.6cm}
\includegraphics[scale=0.63]{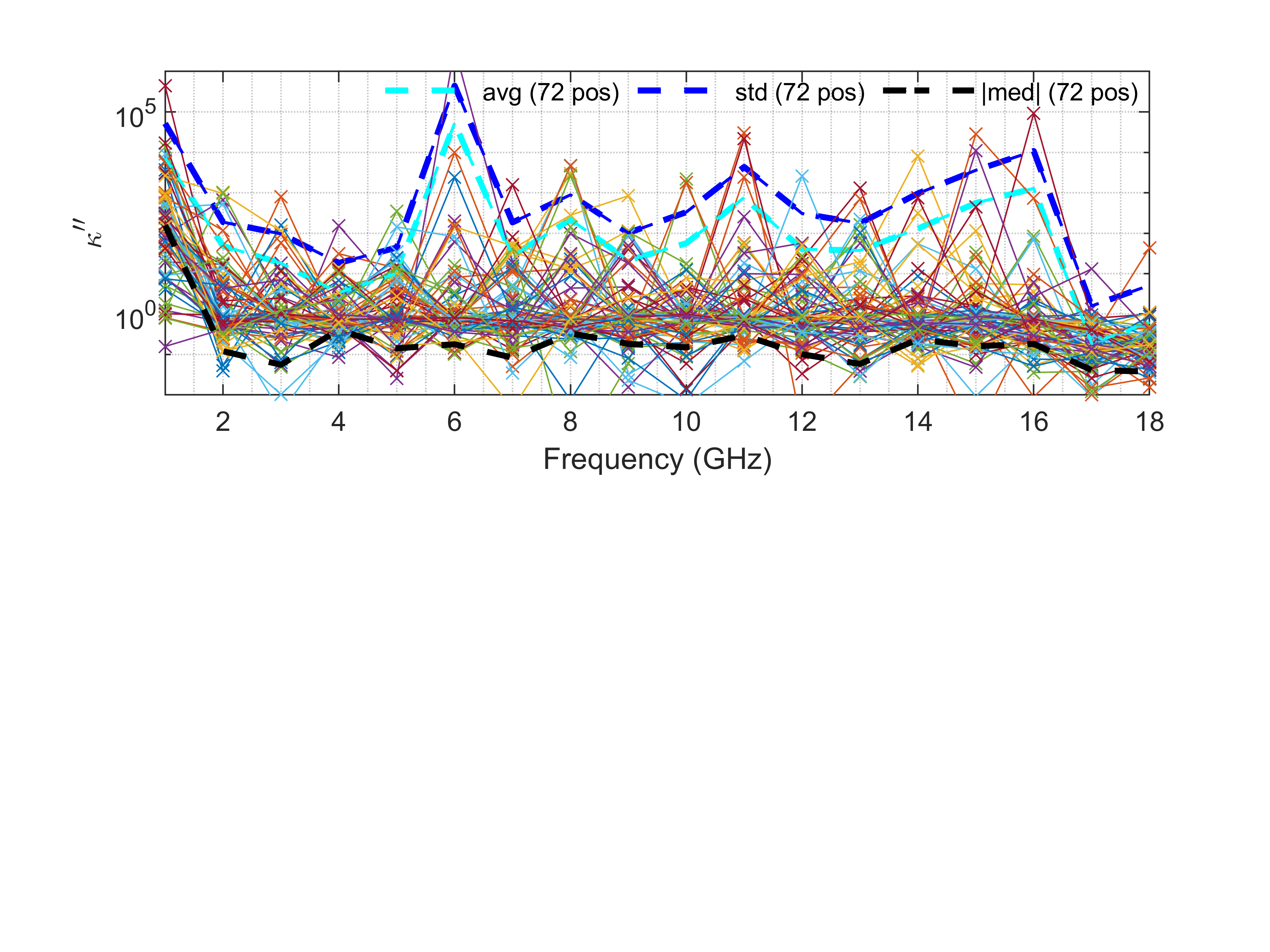}\\ 
\\
\vspace{-5.2cm}\\
\\
(b)\\
\hspace{-0.6cm}
\includegraphics[scale=0.63]{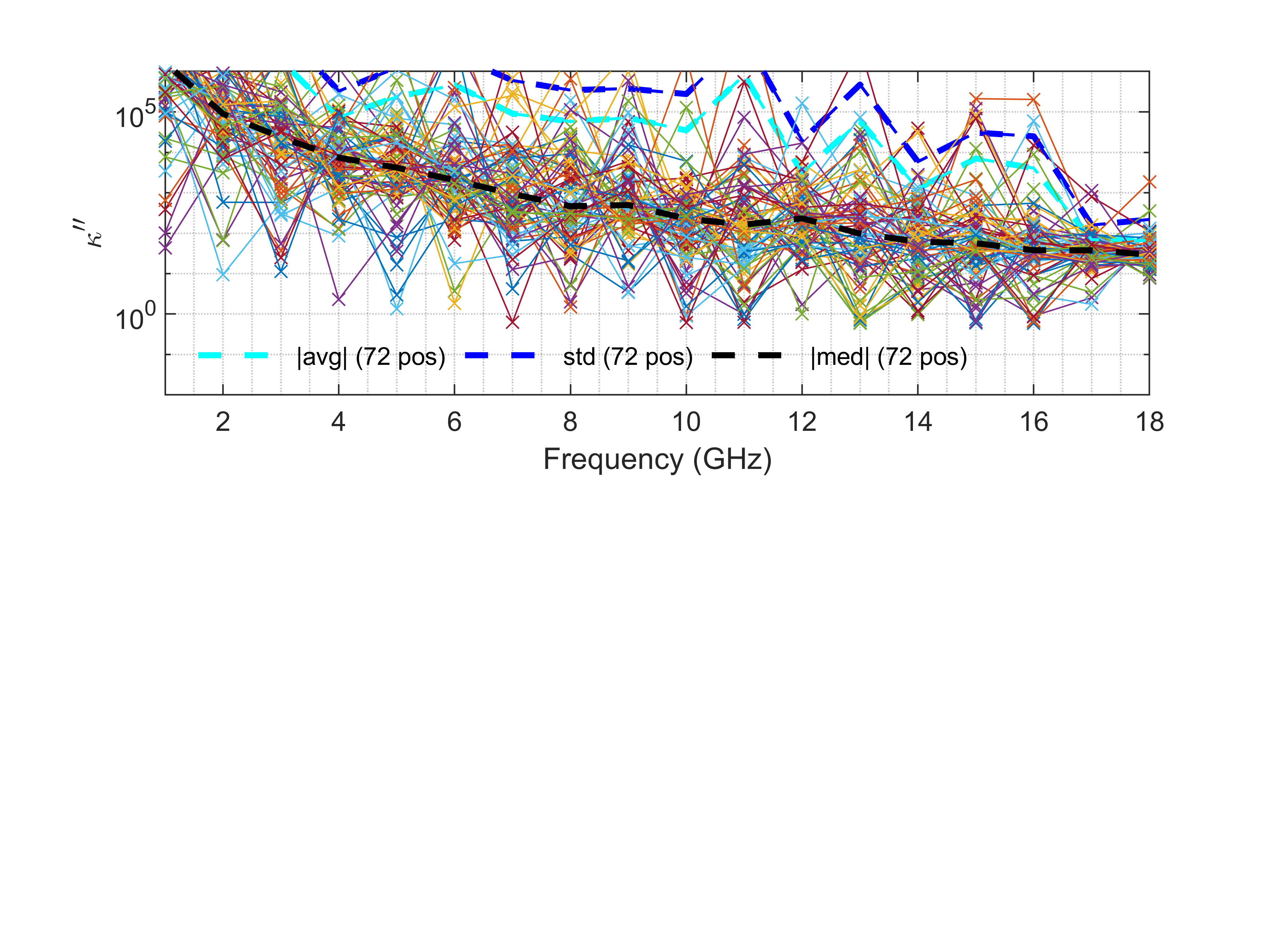}\\ 
\\
\vspace{-5.2cm}\\
\\
(c)\\
\\
\end{tabular}
\end{center}
{
\caption{\label{fig:oddmoments_f1to18GHz}
\small
{
Same as Fig. \ref{fig:evenmoments_f1to18GHz} but for odd-order moments.}
}
}
\end{figure}

Fig. \ref{fig:oddmoments_f1to18GHz}(a) shows the frequency characteristics $(\lambda_m(f)/\lambda_0(f))^{1/m}$ for $m=1,3,5$. 
Compared to those for even-order moments, considerably larger variations occur again across secondary tune states;
similarly for ${\kappa^{\prime\prime}}(f)$ in Fig. \ref{fig:oddmoments_f1to18GHz}(b). 
At high $f$, the convergence of ${\kappa^{\prime\prime}}(f)$ to zero is slower.

Fig. \ref{fig:oddmoments_f1to18GHz}(c) indicates that the ACV method predicts relatively large magnitudes of $\langle\kappa^{\prime\prime}(f)\rangle_{N_t}$ at any frequency, and exhibits a slower decrease with $f$ compared to the stir sweep method. 
The difference is attributable to the values of $\lambda_5(f)/\lambda_0(f)$, which increase more slowly in the ACV method because of the less accurate estimation of the ACF phase angle derivative $\phi^{(5)}_{\rho_E}(0)$ at lower frequencies. 
A similarly slower decrease is found for $\langle \kappa^\prime(f) \rangle_{N_t}$ in Fig. \ref{fig:evenmoments_f1to18GHz}(c), owing to larger values and fluctuations of $\lambda_4(f)/\lambda_0(f)$ traceable to $a^{(4)}_{\rho_E}(0)$ at lower $f$.

\subsection{Dependence on Tune States and Effect of FD Debiasing}
\begin{figure}[htb] 
\begin{center}
\begin{tabular}{c}
\hspace{-0.6cm}
\includegraphics[scale=0.63]{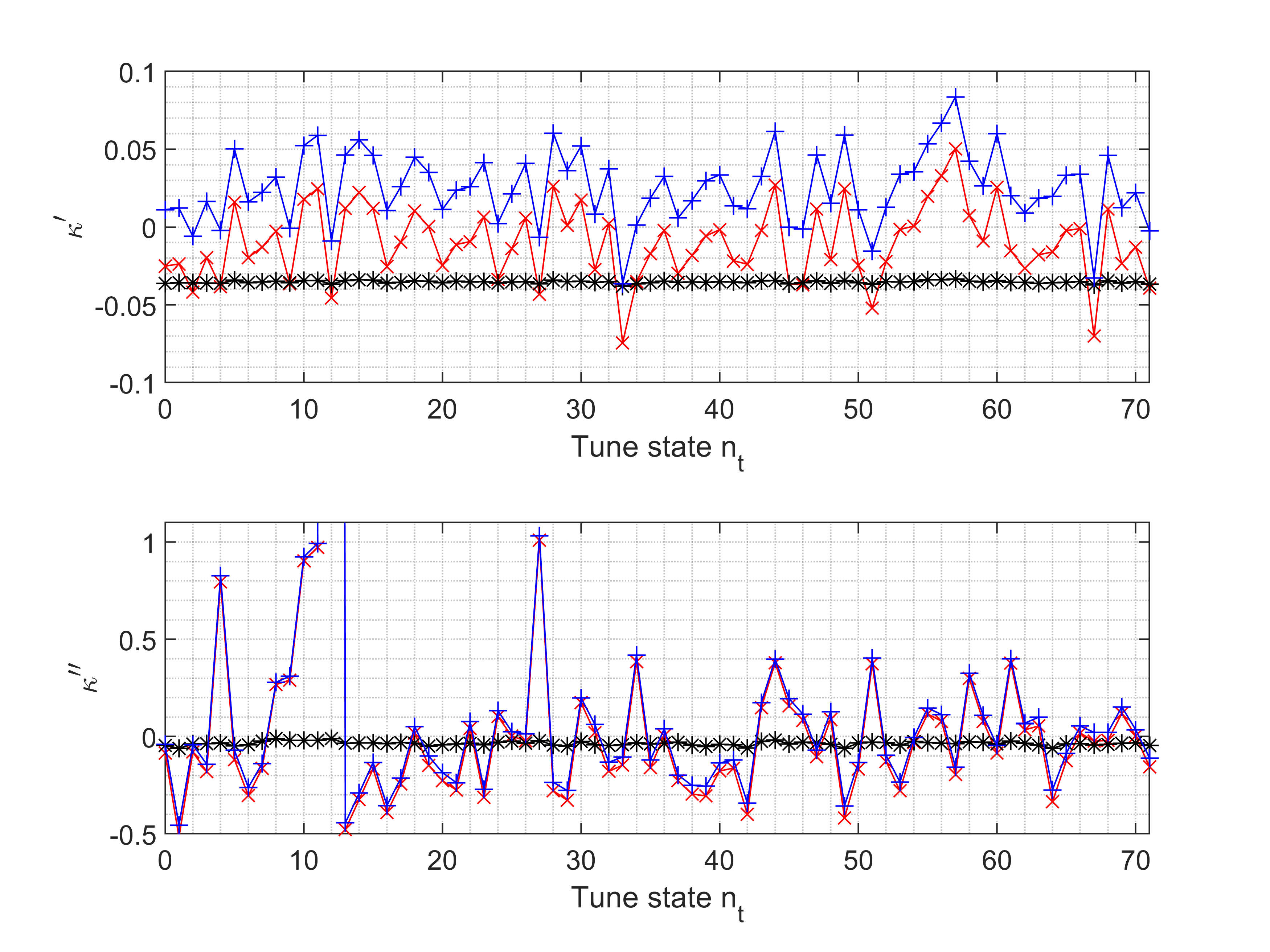}\\ 
\\
\end{tabular}
\end{center}
{
\caption{\label{fig:kappas_pos_ and_f1to18GHz}
\small
{
${\kappa^\prime}$ and ${\kappa^{\prime\prime}}$ across $72$ secondary tune states at $f=18$ GHz: raw data (red), estimated FD bias (black), FD debiased (blue).}
}
}
\end{figure}

Fig. \ref{fig:kappas_pos_ and_f1to18GHz} compares fluctuations of $\kappa^{\prime(\prime)}$ across all tune states before and after FD debiasing at $f=18$ GHz, using the stir sweep FD method. While fewer isolated tune state maintain a negative $\kappa^\prime$, the vast majority incur a positive shift to small values as a result of FD debiasing, as expected in view of the observed LF-to-HF level drop in the ASDF for $\kappa^\prime > 0$ \cite[eq. (43)]{arnaACFSDF_pt1}. By contrast, FD debiasing has a significantly smaller effect on $\kappa^{\prime\prime}$ across all secondary tune states. 

\section{Conclusion}
Sampling, FD, noise, interference, and understirring/line-of-sight coupling in a continuously and uniformly stirred field all affect its spectral and correlation characteristics. These were analyzed theoretically and demonstrated experimentally in this article. Absorption is another major contributor, which will be investigated in future work.

Direct FD of a sampled stir sweep trace, or its complex ACF using (\ref{eq:lambda0_complexACFmethod})--(\ref{eq:phi5_complexACFmethod}), were shown to yield comparable numerical estimates of spectral moments at high excitation frequencies, provided the FD bias is compensated for; cf. Tbl. \ref{tbl:sweep_vs_corr}. The former technique can be implemented as an online (real-time) method, while the latter requires batch processing after all data samples have been acquired.
To leading order, the spectral moments, (\ref{eq:estlambdapeven_cont_fd_approx}) and (\ref{eq:def_lambdaodd_discr_approx_explicit}),
as well as the scaled spectral kurtoses,
(\ref{eq:bias_kappap_fd_cont}) and (\ref{eq:bias_kappapp_fd_cont}), exhibit a negative quadratic bias with respect to the normalized sampling resolution, $\Delta \tau/\beta^{\prime(\prime)}$, that can be compensated for. 
Odd-order spectral moments $\lambda^{\prime\prime}_{2i+1}$ and $\kappa^{\prime\prime}(f)$ involve unequal orders of FD and exhibit considerably larger spread for individual stir traces than even-order $\lambda^\prime_{2i}$ and $\kappa^\prime(f)$.

Noise, interference and understirring were shown to have a major effect on the spectral kurtosis of ideal stirred continuous fields ($\kappa^\prime = 0$) via (\ref{eq:kappap_E0plusN}), (\ref{eq:kappap_E0plusEh}) and (\ref{eq:kappa_E0plusEu}). A rich dependence structure on noise-to-stir and noise-to-interference ratios of levels and bandwidths was found (Figs. \ref{fig:kappap_ifv_SNR} and \ref{fig:John_specmom_ifv_Pin}). 
This demonstrates the high sensitivity of CFs and SDFs on field imperfections, even after FD bias compensation.
The results in Figs. \ref{fig:evenmoments_f1to18GHz}--\ref{fig:kappas_pos_ and_f1to18GHz} illustrate the challenge of accurately estimating the small magnitude and sign of $\kappa^{\prime(\prime)}$. 

Whereas the analysis for second- and higher-order Pad\'{e}-based models is more involved, the simple $[0/1]$ model limits the effects on sampling, noise and interference to a consideration of the parameter $\lambda^\prime_2$ only.

\end{document}